\documentclass[conference]{IEEEtran}
\IEEEoverridecommandlockouts
\usepackage{cite}
\usepackage{amsmath,amssymb,amsfonts}
\usepackage{algorithmic}
\usepackage{algorithm}
\usepackage{graphicx}
\usepackage{textcomp}
\usepackage{xcolor}
\usepackage{flushend}
\usepackage{url}
\usepackage{booktabs}
\usepackage{svg}
\usepackage{tikz}
\def\BibTeX{{\rm B\kern-.05em{\sc i\kern-.025em b}\kern-.08em
    T\kern-.1667em\lower.7ex\hbox{E}\kern-.125emX}}
\begin{document}

\title{Resource-Aware Deployment Optimization for Collaborative Intrusion Detection in Layered Networks}

\author{
André García Gómez\IEEEauthorrefmark{1}, Ines Rieger\IEEEauthorrefmark{2}, Wolfgang Hotwagner\IEEEauthorrefmark{1}, Max Landauer\IEEEauthorrefmark{1},\\ Markus Wurzenberger\IEEEauthorrefmark{1}, Florian Skopik\IEEEauthorrefmark{1}, Edgar Weippl\IEEEauthorrefmark{3}\\
\IEEEauthorblockA{\IEEEauthorrefmark{1}AIT Austrian Institute of Technology, Vienna, Austria 
Email: firstname.lastname@ait.ac.at} 
\IEEEauthorblockA{\IEEEauthorrefmark{2} Airbus, Germany
Email: firstname.lastname@airbus.com}
\IEEEauthorblockA{\IEEEauthorrefmark{3} University of Vienna, Vienna, Austria
Email: firstname.lastname@univie.ac.at}
}

\maketitle

\begin{abstract}
Collaborative Intrusion Detection Systems (CIDS) are increasingly adopted to counter cyberattacks, as their collaborative nature enables them to adapt to diverse scenarios across heterogeneous environments. As distributed critical infrastructure operates in rapidly evolving environments, such as drones in both civil and military domains, there is a growing need for CIDS architectures that can flexibly accommodate these dynamic changes. In this study, we propose a novel CIDS framework designed for easy deployment across diverse distributed environments. The framework dynamically optimizes detector allocation per node based on available resources and data types, enabling rapid adaptation to new operational scenarios with minimal computational overhead. We first conducted a comprehensive literature review to identify key characteristics of existing CIDS architectures. Based on these insights and real-world use cases, we developed our CIDS framework, which we evaluated using several distributed datasets that feature different attack chains and network topologies. Notably, we introduce a public dataset based on a realistic cyberattack targeting a ground drone aimed at sabotaging critical infrastructure. Experimental results demonstrate that the proposed CIDS framework can achieve adaptive, efficient intrusion detection in distributed settings, automatically reconfiguring detectors to maintain an optimal configuration, without requiring heavy computation, since all experiments were conducted on edge devices.
\end{abstract}


%/\tableofcontents
\section{Introduction}
Cyber attacks are becoming increasingly significant in both civilian and military infrastructure \cite{md_haris_uddin_sharif_literature_2022, carlo_cyber_2024}. As the Internet of Things (IoT) infrastructure expands with home devices, industrial sensor devices, and the proliferation of unmanned aerial vehicles (UAVs) and ground vehicles (UGVs), the attack surface has increased. In 2015, Al-Fuqaha et al.~\cite{al-fuqaha_internet_2015} projected 50 billion connected devices by 2020. This growth has been accompanied by an increase in vulnerabilities in the IoT infrastructure and new comprehensive threat models that include various attack vectors targeting end-user devices \cite{stellios_survey_2018}.

Recent real-world conflicts, such as the Ukraine-Russia war, where the use and technological sophistication of drones have risen sharply, as outlined by the News Strategy Center\footnote{Read 17-10-2025, https://newstrategycenter.ro/wp-content/uploads/2024/02/Lessons-Learned-from-the-War-in-Ukraine.-The-impact-of-Drones-2.pdf} also underscore this upward trend in complexity and attack surface. A similar focus has been observed in civilian IoT domains, where stealth attacks on phone networks \cite{andreolini_collaborative_2015} or cyber threats in smart cities were analyzed \cite{alrashdi_ad-iot_2019}.

To defend such distributed environments, Collaborative Intrusion Detection Systems (CIDS) have been developed, where multiple Intrusion Detection Systems (IDS) collaborate to detect threats. However, many existing systems feature static architectures that may struggle to adapt to rapid network changes. For example, reflecting on the Ukraine-Russia war, where kinetic attacks have been proven to be highly effective \cite{aviv_russian-ukraine_2023}, automatic reconfiguration of the systems is required when certain nodes are compromised or need disconnection to avoid detection on the battlefield.

As noted by \cite{vasilomanolakis_taxonomy_2015}, classic CIDS often rely on hierarchical topologies that risk single points of failure, or use peer-to-peer techniques that impose high computational overhead. Our research focuses on exploring a more flexible approach, allowing CIDS to adapt to hardware requirements, similar to the features discussed in \cite{salehie_self-adaptive_2009}. Although self-configuration methods such as \cite{chiba_automatic_2022}, which propose an automated hyperparameter tuning technique for deep neural network IDS, and \cite{beck_semi-supervised_2024}, which describe an auto-configuration tool for anomaly-based IDS, have been explored, these approaches lack cross-node coordination. They operate within individual detectors, while our proposed method can automatically identify which IDSs are active and coordinate their deployment across systems, remaining scalable even in complex scenarios involving a large number of distributed nodes. 

To develop a CIDS framework capable of supporting such adaptability, we began by performing a literature review to identify the main characteristics of existing architectures and to examine different operational scenarios of IoT infrastructure. To support the evaluation in realistic scenarios, we also contributed a public dataset based on a cyber attack targeting a UGV connected to other systems tasked with monitoring a hard-to-access area. To show that the proposed CIDS framework can flexibly adapt to various deployment setups and be deployed on real data within a simulated real-world environment, we structure our experiments around the following research questions (RQ):
\begin{itemize}
    \item \textbf{RQ1: How effectively does the CIDS framework operate across different edge device architectures?} We construct several scenarios with varying network scales, hardware specifications, and IDS configurations. This setup evaluates performance and configuration speed since the optimization must be completed in a reasonable time to ensure practical deployment feasibility.
    \item \textbf{RQ2: How well does the proposed framework perform when deployed in a realistic multi-domain scenario?} We emulate a realistic environment using real IDS nodes that are interconnected and multi-domain datasets, including our novel UGV attack dataset, to validate operational robustness.
\end{itemize}

The main contributions of this paper are:
\begin{itemize}
    \item A systematic and detailed review of existing CIDS architectures including an explanation of the different CIDS components. We propose a refined classification scheme to categorize CIDS that better captures the functional and topological diversity in modern distributed environments.
    \item A novel, adaptive CIDS framework that offers flexible automated optimization, prioritizing the deployment of detectors across heterogeneous nodes and system architectures, while accounting for varying resource constraints of the underlying hardware.
    \item A publicly available dataset was created that shows a multi-stage cyberattack executed on an UGV to support realistic evaluation. The dataset can be found here \cite{hotwagner_2026_18482763}.
    \item Empirical validation of the framework using both synthetic datasets and real IDS deployments in a distributed setting, including the new UGV dataset, to measure the effectiveness of the proposed CIDS framework.
\end{itemize}

The paper is structured as follows: Section \ref{sec:related_work} reviews related work and defines the core CIDS components. Section \ref{sec:cids_analysis} presents our classification and analysis of existing architectures. Section \ref{sec:propose_architecture} details our proposed framework and optimization algorithms. Section \ref{sec:robot_dog} describes the UGV dataset created for this paper. Section \ref{sec:experimental_setup} outlines the setup used for the empirical experiments. Section \ref{sec:results} presents the collected results, and Section \ref{sec:discussion} analyzes and interprets these findings. Section \ref{sec:threat} analyzes different threats to the validity of our experiments. Lastly, Section \ref{sec:conclusion} summarizes the main conclusions and suggests directions for future research. The code is available here\footnote{https://github.com/ait-aecid/cids-evaluation/tree/main}.

\section{Related Work}
\label{sec:related_work}

\begin{figure}[h]
    \centering
    \includegraphics[width=7cm]{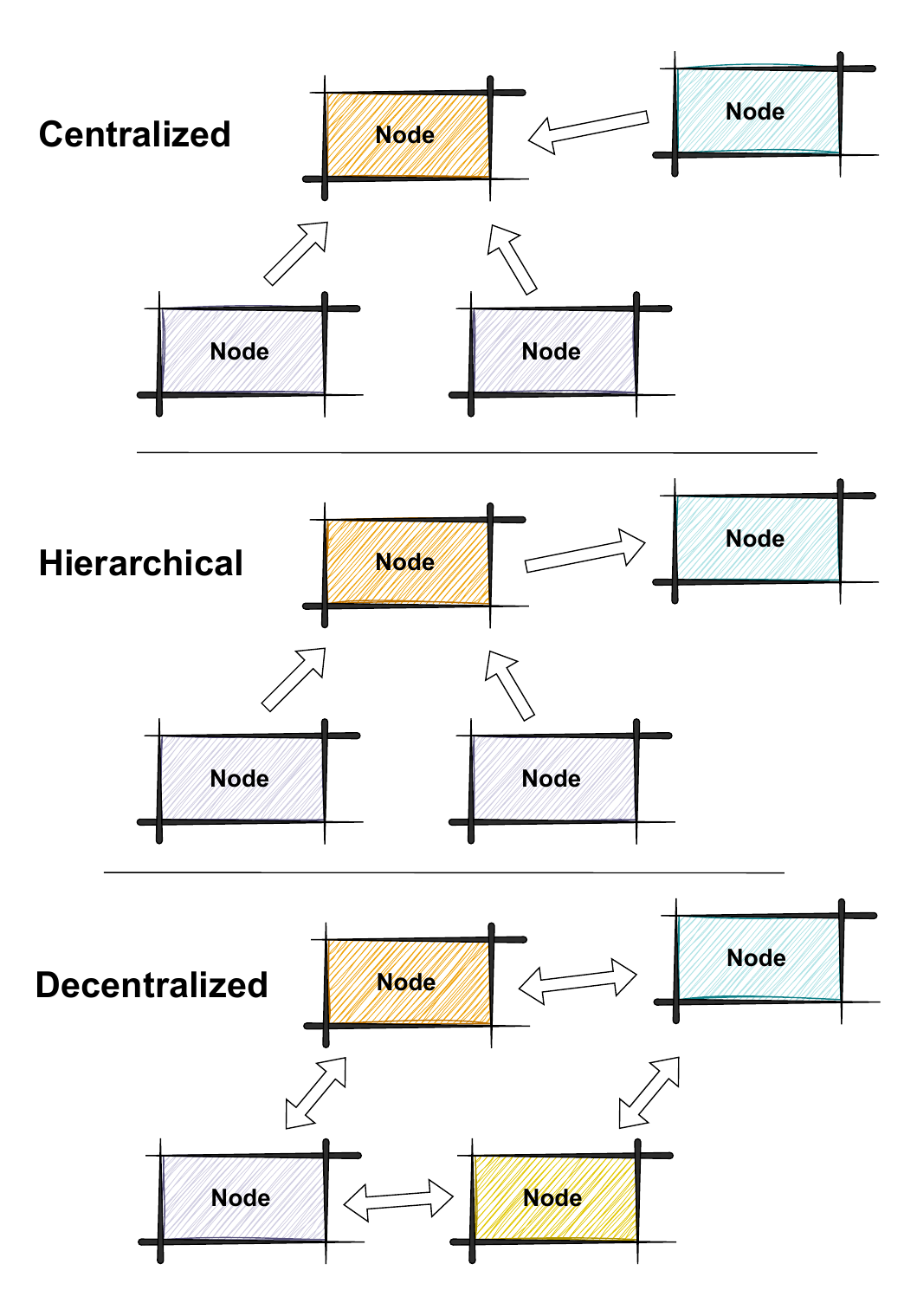}
    \caption{CIDS nodes distributions architectural patterns based on \cite{vasilomanolakis_taxonomy_2015, zhou_survey_2010, garcia_gomez_collaborative_2026}.}
    \label{fig:cids_types}
\end{figure}

CIDS are intricate systems comprising various components. This section will emphasize its main components based on previous research and surveys \cite{vasilomanolakis_taxonomy_2015, zhou_survey_2010, garcia_gomez_collaborative_2026}:
\begin{itemize}
    \item \textbf{Intrusion Detection Methods}: These are the primary detection algorithms utilized within the system. As highlighted in \cite{garcia_gomez_collaborative_2026}, multiple algorithms may require different types of data input, such as network traffic, log data, or time series. A few IDS methods can be found in \cite{garcia_gomez_collaborative_2026, abdulganiyu_systematic_2023, lansky_deep_2021}.
    \item \textbf{Manager}: Managing distributed systems involves handling nodes and deploying system components.
    \item \textbf{Alert Aggregation}: Techniques for combining the various alerts generated by multiple detection methods \cite{landauer_dealing_2022}. Nevertheless, it may also be found in the literature under alternative names like Manager \cite{wu_collaborative_2003} or Alert Clustering \cite{cuppens_alert_2002}.
    \item \textbf{Message Queue}: The message queue in CIDS enables node interaction, with each implementation using different protocols and technologies. RFC4765 (IDMEF) \cite{debar_rfc_2007} is frequently cited as the standard format for the exchange of information between IDS, but its application in the industry may be limited.
    \item \textbf{Security and Trust}: Distributed systems offer flexibility but also increase potential attack vectors, with each node interaction being a vulnerability point. Attacks might create fake nodes that disrupt operations or collect communication data.
    \begin{itemize}
        \item \textit{Trust layer}: Studies such as \cite{duma_trust-aware_2006, ezelu_membership_2023, fung_robust_2009} reduce this risk by assessing the level of trust of each node relative to others. Other research employs a blockchain approach to authenticate system entities \cite{gurung_cids_2022}.
        \item \textit{Security layer}: Although numerous studies recognize the significance of securing communications, there is little novelty introduced here, as some studies merely refer to the application of private-public keys \cite{zhou_evaluation_2007, jalil_hadi_real-time_2024}.
    \end{itemize}
    \item \textbf{Node Distribution}: Historically, the primary distinction between CIDS has been categorized by dividing them into three distribution groups \cite{vasilomanolakis_taxonomy_2015, zhou_survey_2010, garcia_gomez_collaborative_2026}, as shown in Figure \ref{fig:cids_types}:
    \begin{itemize}
        \item \textit{Centralized}: Alerts are centralized in a single node, simplifying information management \cite{arshad_colide_2019, landauer_AMiner_2023}, but create a single point of failure \cite{vasilomanolakis_taxonomy_2015}.
        \item \textit{Hierarchical}: This architecture surpasses the centralized CIDS by using a hierarchical tree for alert distribution. Despite having single point of failure, a well-deployed structure remains robust and efficient \cite{hardegen_hierarchical_2021}.
        \item \textit{Decentralized}: This approach requires the nodes to engage in peer-to-peer communication, thus eliminating a single point of failure. Ensure redundancy by increasing the number of communications \cite{duma_trust-aware_2006, vasilomanolakis_skipmon_2015}.
    \end{itemize}
\end{itemize}

\section{CIDS analysis}
\label{sec:cids_analysis}

Before designing our CIDS, we thoroughly analyze the literature. Current systems are categorized into centralized, distributed, and decentralized nodes \cite{vasilomanolakis_taxonomy_2015, zhou_survey_2010}, but this is too broad. It lacks detail on distribution of alert correlation, security complexity, or training considerations. We will offer an alternative view of previous research, highlighting aspects for our design.

\subsection{CIDS publications}
Our analysis of CIDS starts by reviewing surveys \cite{vasilomanolakis_taxonomy_2015, zhou_survey_2010, garcia_gomez_collaborative_2026} and using a snowball method on Google Scholar. This paper does not aim to be a comprehensive survey, but highlights key characteristics and trends for our design proposal. Table \ref{tab:new_cids} shows the different publications. Due to fewer publications in this field, we extended the timeframe for a more representative selection (2000-2025). The following is an explanation of the table fields.
\begin{table}
    \caption{CIDS publications organized by year. Older publications are divided from the rest with a horizontal line.}
    \centering
    \begin{tabular}{c|c|c|c}
        \textbf{Publication} & \textbf{Corr. units}  & \textbf{Security} &  \textbf{Training} \\ \hline \hline
        Javeed et al. 2024 \cite{javeed_federated_2024} & None  & Low  & Federated \\
        Tlili et al. 2024 \cite{tlili_exhaustive_2024} & Single  & Low  & Centralize \\
        Sadia et al. 2024 \cite{sadia_intrusion_2024} & None  & Low  & Centralize \\
        Turukmane et al. 2024 \cite{turukmane_m-multisvm_2024} & None  & Low  & Centralize \\
        Jalil Hadi et al. 2024 \cite{jalil_hadi_real-time_2024} & Strong Dist.  & Medium  & Centralize \\
        Landauer et al. 2023 \cite{landauer_AMiner_2023} & Single  & Low  & Centralize \\
        Javed et al. 2023 \cite{javed_prism_2023} & Single  & Low  & Centralize \\
        Ezelu et al. 2023 \cite{ezelu_membership_2023} & Hierarchy  & Medium  & Centralize \\
        Hnamte et al. 2023 \cite{hnamte_dcnnbilstm_2023} & None  & Low  & Centralize \\
        Gurung et al. 2022 \cite{gurung_cids_2022} & Strong Dist. & High  & None \\
        Alsaadi et al. 2022 \cite{alsaadi_adapting_2022} & None & Low  & Centralize \\
        Li et al. 2022 \cite{li_federated_2022} & None & Low  & Federated \\
        Hardegen et al. 2021 \cite{hardegen_hierarchical_2021} & Hierarchy & Low  & Centralize \\
        Arshad et al. 2019 \cite{arshad_colide_2019} & Single & Low  & None \\
        Nguyen et al. 2019 \cite{nguyen_diot_2019} & None & Low  & Federated \\ \hline
        Vasilomanolakis et al. 2015 \cite{vasilomanolakis_skipmon_2015} & Weak Dist.  & Low  & None \\
        Morais et al. 2014 \cite{morais_distributed_2014} & Strong Dist.  & Low  & None \\
        Modi et al. 2013 \cite{modi_novel_2013} & Single  & Low  & Centralize \\
        Xiang et al. 2012 \cite{xiang_taxonomy_2012} & None & Low  & Centralize \\
        Kholidy et al. 2012 \cite{kholidy_cids_2012} & Hierarchy & Low  & Centralize \\
        Gul et al. 2011 \cite{gul_distributed_2011} & None & Low  & None\\
        Roschke et al. 2010 \cite{roschke_advanced_2010} & Single & Low  & None\\
        Vieira et al. 2010 \cite{vieira_intrusion_2010} & Strong Dist. & Low  & Centralize\\
        Fung et al. 2009 \cite{fung_robust_2009} & Strong Dist. & Medium  & None\\
        Zhou et al. 2007 \cite{zhou_evaluation_2007} & Strong Dist. & Medium  & None\\
        Duma et al. 2006 \cite{duma_trust-aware_2006} & Strong Dist. & Medium  & None\\
        Dasgupta et al. 2005 \cite{dasgupta_cids_2005} & Strong Dist. & Medium  & Centralize\\
        Min Cai et al. 2005 \cite{min_cai_collaborative_2005} & Weak Dist. & Medium  & None\\
        Locasto et al. 2005 \cite{locasto_towards_2005} & Weak Dist. & Low  & None\\
        Wu et al. 2003 \cite{wu_collaborative_2003} & Single & Low  & None\\
        Janakiraman et al. 2003 \cite{janakiraman_indra_2003} & Weak Dist. & High & None\\
        Cuppens et al. 2002 \cite{cuppens_alert_2002} & Single & Low & None\\
    \end{tabular}
    \label{tab:new_cids}
\end{table}

\subsection{CIDS classification}
Figures \ref{fig:sec_dist} and \ref{fig:train_dist} present the proposed classification for different CIDS. Detailed explanations of the categories follow below. All referenced publications are listed in Table \ref{tab:new_cids}.

\subsubsection{Node Distribution}
Previous classifications were limited to communication between different nodes. Our approach improves this classification by including additional data. We differentiate based on workload and communication levels between detectors and alert aggregations. The communication patterns between detectors and alert aggregators vary between architectures, influencing the workload assigned to each alert aggregator. Figures \ref{fig:sec_dist} and \ref{fig:train_dist} represent the workload using three distinct color codes:
\begin{itemize}
    \item \textit{Low workload (Green)}: Most alert aggregators manage alerts from a relatively limited number of detectors.
    \item \textit{Medium workload (Purple)}: Each alert aggregator tackles alerts from a substantial number of detectors.
    \item \textit{High workload (Orange)}: All aggregators handle the alerts from every detector.
\end{itemize}
A comparable categorization of workload is presented in \cite{zhou_survey_2010}. The specific classifications for node distributions are as follows:
\begin{itemize}
    \item \textbf{None}: This set of publications does not incorporate an alert aggregator, thus lacking a designated color code for workload.
    \item \textbf{Single}: A single alert aggregator is utilized in this publication to handle all system alerts, assigned an orange color code.
    \item \textbf{Hierarchy}: Detectors are organized into groups, with each group channeling alerts to a specific alert aggregator or higher-tier ones in a hierarchical structure. As lower-tier aggregators manage alerts only for a limited set of detectors, this group is categorized with a green code.
    \item \textbf{Weak distribution}: Nodes are partitioned into groups, and each node houses a detector and an alert aggregator. Nodes within a group engage in peer-to-peer communication, with some nodes collecting alerts from various groups. This setup is labeled with a purple color code, indicating a high workload for aggregators, though not all handle every alert.
    \item \textbf{Strong distribution}: The entire system operates on a full peer-to-peer network, with each node equipped with a detector and an alert aggregator. All aggregators are responsible for receiving alerts from every detector, marked by an orange color code.
\end{itemize}

\subsubsection{Security Complexity}

We incorporate a categorization system for security and trust. Security is responsible for ensuring secure communication between nodes, whereas trust involves evaluating and classifying each node's trustworthiness within the system. Diverse methods and technologies are introduced in various publications, complicating empirical evaluation. Though a detailed comparison of different performances against various attacks is beyond the scope of this paper, we opt to categorize the CIDS based on the computational complexity inherent in these security and trust methods.
\begin{itemize}
    \item \textbf{Low}: Papers in this group feature minimal or nonexistent approaches to trust and security.
    \item \textbf{Medium}: These papers present a significant method for security and trust, which forms a key aspect of its contributions.
    \item \textbf{High}: The approach to security and trust requires the use of an external infrastructure, such as a blockchain.
\end{itemize}

\subsubsection{Training setup}

Although most publications report their detection performance, not all examine the details of the method setup or training processes. We categorize these into the following groups: 
\begin{itemize} 
    \item \textbf{None}: The paper does not involve any training. Instead, pre-configured external detectors are used, or empirical experiments were not done. 
    \item \textbf{Centralized}: Detectors require data for training, each node independently trains its detectors using the data it collects, and no communication occurs between nodes during training. 
    \item \textbf{Federated}: A federated training approach is employed, similar to the one described in \cite{mcmahan_communication-efficient_2023}.
\end{itemize}

\begin{figure}[h]
    \centering
    \includegraphics[width=9.5cm]{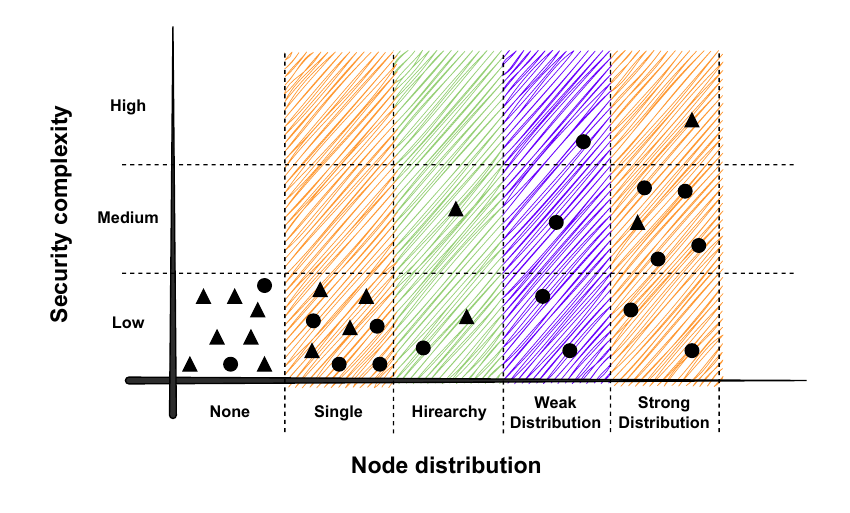}
    \caption{The comparison of various CIDS publications is made by examining Security complexity and Node distribution. Publications from the last decade  are depicted as triangles, while older ones are shown as circles.}
    \label{fig:sec_dist}
\end{figure}

\begin{figure}[h]
    \centering
    \includegraphics[width=9.5cm]{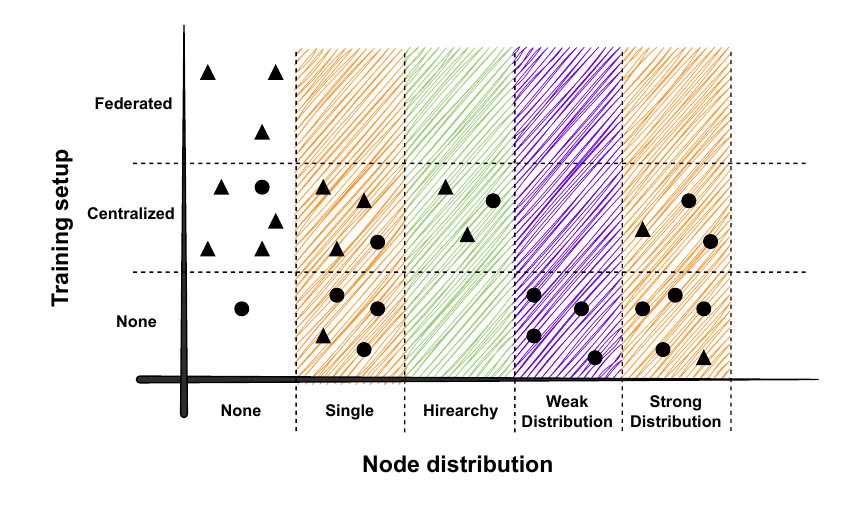}
    \caption{The comparison of various CIDS publications is made by examining Training setup and Node distribution. Publications from the last decade  are depicted as triangles, while older ones are shown as circles.}
    \label{fig:train_dist}
\end{figure}

\subsection{Main Insights Gathered}
Based on the various previously mentioned publications and the examination of Figures \ref{fig:sec_dist} and \ref{fig:train_dist}, we present the following key points:\begin{itemize}
    \item \textbf{Single points of failure}: A single point of failure, like one alert aggregator, poses a risk. A strong distribution shares this load but adds complexity (Figure \ref{fig:sec_dist}). In some cases, these risks are minimal. AMiner \cite{landauer_AMiner_2023} is monolithic with one alert aggregator, meaning its failure is only caused by a total system failure as it is not on a separate node. Conversely, a distributed system offers redundancy but increases complexity and potential new vulnerabilities.
    \item \textbf{Security vs Node distribution}: Figure \ref{fig:sec_dist} shows the distribution of publications addressing security complexity and node distribution. Studies using multiple alert aggregators focus more on security, but most do not. Although crucial, security and trust often get overlooked in system design, despite their importance early in the process \cite{shostack_threat_2014}.
    \item \textbf{Training vs Node distribution}: Publications using various node distributions often neglect the training differences of IDS. As noted in \cite{garcia_gomez_collaborative_2026}, not all IDS methods need training data, but all require data for performance evaluation. Unequal data distribution among nodes may lead to insufficient training data for some, which can be shared within an organization or when privacy is not a concern. Federated learning \cite{mcmahan_communication-efficient_2023} addresses this but requires a central node, risking a single point of failure. To solve this, some propose peer-to-peer federated learning \cite{wink_approach_2021, martinez_beltran_decentralized_2023}. Alternatively, \cite{dautov_latency-aware_2024} adds latency awareness for robustness in the training process.
    \item \textbf{Static architectures}: COLIDE \cite{arshad_colide_2019} and PRISM \cite{javed_prism_2023} suggest architectures for specific scenarios, limiting versatility. We advocate for an architecture independent of the IDS or data, and facilitating deployment in rapidly changing IoT environments in civil and military sectors. To extend system lifecycles, systems should auto-optimize based on hardware and ongoing changes. While some methods offer flexibility, this area remains underexplored.
    \item \textbf{External dependencies}: Certain publications, like \cite{gurung_cids_2022} using smart contracts among nodes and \cite{janakiraman_indra_2003} with Web of Trust node management, show that successful deployment relies on third-party infrastructure. Similar challenges in IDS include using third-party LLMs like ChatGPT for anomaly IDS \cite{garcia_gomez_collaborative_2026}, introducing risks if services are discontinued or unavailable.
    \item \textbf{Historical trend}: Based on the publications previously shown in Table \ref{tab:new_cids}, a correlation between the papers is observable. Papers from the current decade generally emphasize simpler architectures and incorporate IDS methods like machine learning techniques that require training steps. In contrast, older papers feature more complex architectures, such as peer-to-peer methods, with IDS approaches that often do not require training. This trend may be coincidental but it follows the rise of deep learning methods in recent years, suggesting that while older studies may have concentrated more on architectural distribution, recent ones have shifted focus towards detection methods. For context, one of the earliest deep learning methods in log anomaly IDS is Deeplog (2017) \cite{noauthor_deeplog_nodate}, and the federated learning paper \cite{mcmahan_communication-efficient_2023} is from 2023.
 \end{itemize}

\section{Proposed CIDS framework}
\label{sec:propose_architecture}

Our CIDS framework was developed by combining the main insights from Section \ref{sec:cids_analysis} with the lessons learned from several IoT use cases described in \cite{wurzenberger_newsroom_2024}. In this section, we first provide an overall description of the framework, followed by a detailed explanation of the deployment optimization.

\subsection{Overall Framework}

\begin{figure}[h]
    \centering
    \includegraphics[width=9cm]{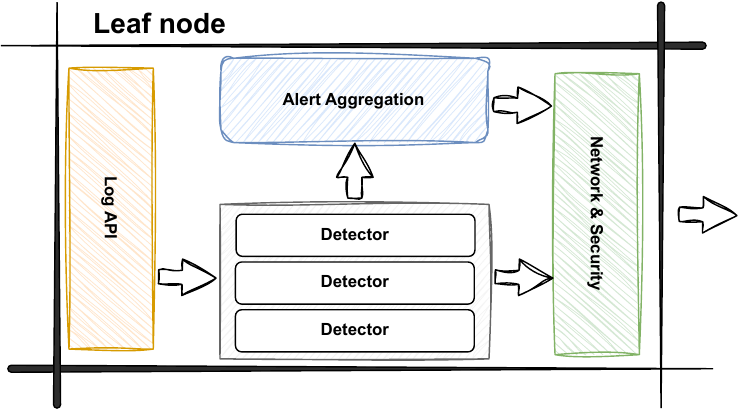}
    \caption{Internal architecture of a Leaf Node.}
    \label{fig:leaf_node}
\end{figure}

\begin{figure}[h]
    \centering
    \includegraphics[width=9cm]{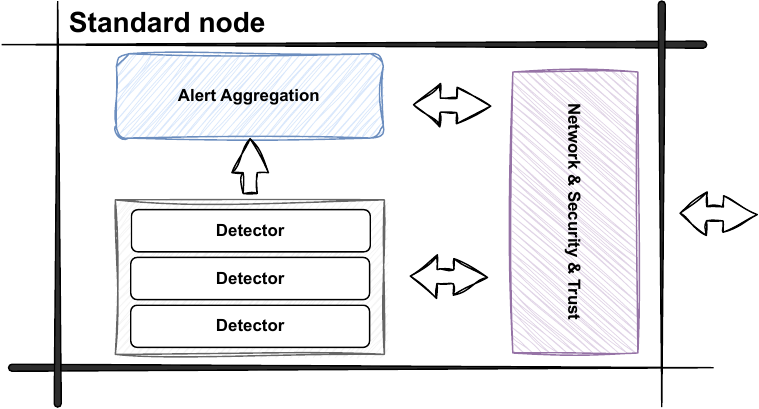}
    \caption{Internal architecture of a Standard Node.}
    \label{fig:main_node}
\end{figure}

The framework design allows deployment across various hardware configurations, with each node incorporating an alert aggregator and multiple detectors. This arrangement effectively converts each node into a sub-CIDS, featuring a singular alert aggregator distribution, thereby granting a certain level of autonomy to each node. Despite introducing a potential single point of failure, as previously noted, the node's architecture remains monolithic. Thus, similar to the rationale provided for AMiner, a considerable failure is needed to render the entire node offline and disrupt the alert aggregator. The main goal is to ensure that, even if communication with higher-level nodes is interrupted or delayed, individual nodes can still generate alerts with some precision and decide on implementing countermeasures. We classify the nodes into two categories:
\begin{itemize}
    \item \textbf{Leaf Node}: As foundational components within the CIDS, the Leaf nodes are located on edge computers and have the role of transmitting system data to the larger network (Figure \ref{fig:leaf_node}).
    \item \textbf{Standard Nodes}: Functioning as higher-level nodes within the CIDS, standard nodes collect system data from various Leaf nodes and other alerts across the network (Figure \ref{fig:main_node}). They execute their local detection processes and reconcile the received alerts with those generated by their detectors. Due to their dual role in data reception and transmission, these nodes include both trust and security layers, in contrast to Leaf nodes, which only have a security layer due to their exclusive function of data transmission.
\end{itemize}

\begin{figure}[h]
    \centering
    \includegraphics[width=9cm]{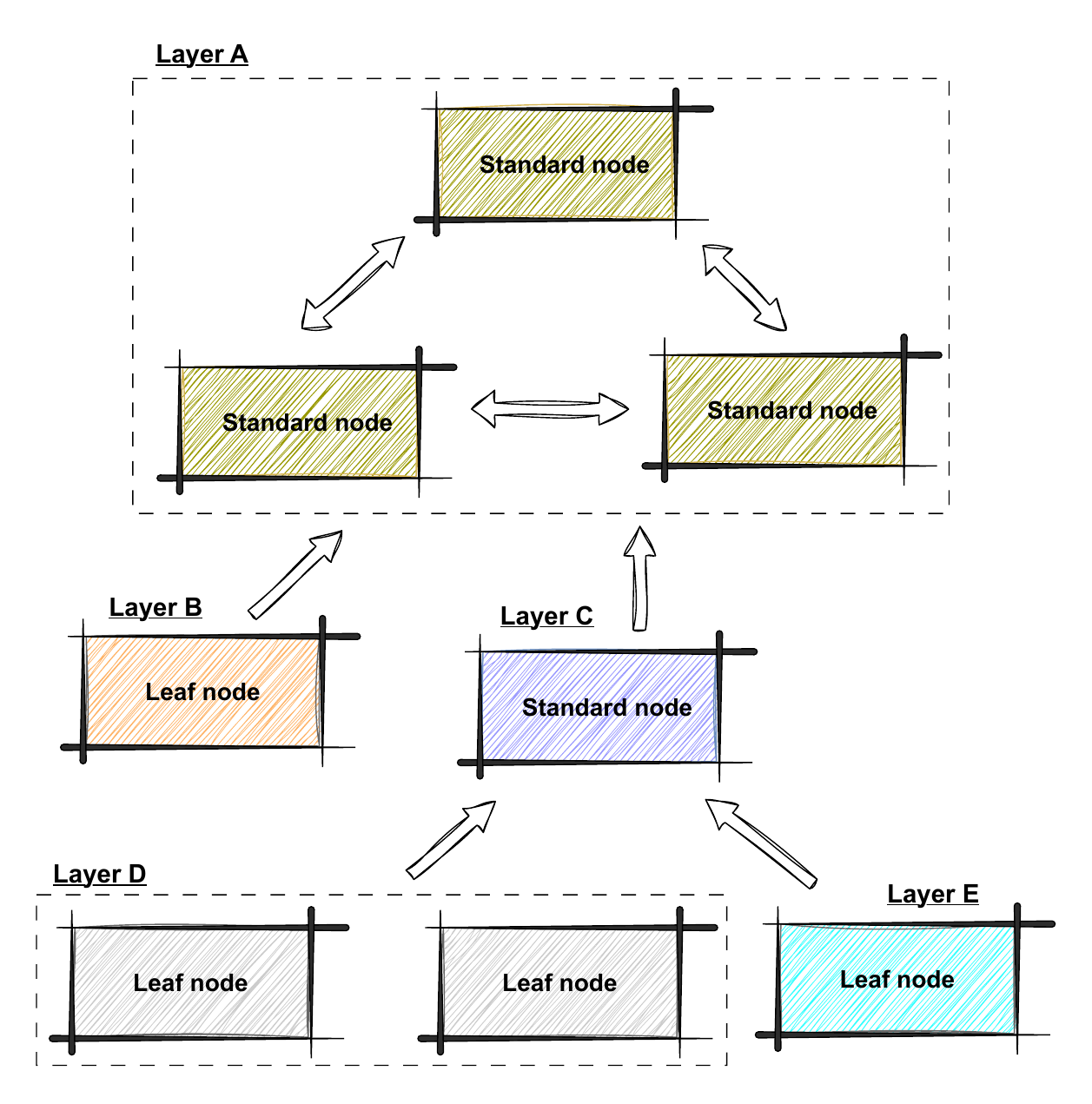}
    \caption{Example of a CIDS architecture based on our propose framework.}
    \label{fig:example_cids}
\end{figure}

The node distribution generally falls between hierarchical and weak distribution, allowing it to adjust between the two based on hardware constraints. Figure \ref{fig:example_cids} illustrates how the infrastructure might appear in a practical setting. Here, nodes are organized into layers, which we will describe in more detail later in the text. The arrows indicate the direction of data flow, with leaf nodes transmitting data to the standard nodes, and all alerts following the arrow's direction within the system. Although the interlayer structure exhibits a hierarchical architecture, the intralayer nodes can be interconnected, minimizing the risk of a single point of failure (Layer A). Should one of these nodes fail, the system continues to function. Each node operates as an independent unit and if only one node is in the layer (Layer C), a single point of failure remains possible. Although the framework offers tools to mitigate this, the infrastructure design and the number of hardware units are also crucial factors. Although we have incorporated trust and security layers in Figures \ref{fig:leaf_node} and \ref{fig:main_node}, the specifics of their operation as the training process are beyond the scope of this paper, as they may vary depending on the scenario or use case.

\subsection{Deployment optimization}

As previously stated, our primary goal is to reduce alert times. To accomplish this, we plan to strategically place each detector at different nodes by using the node optimizer. An initial thought might be to relocate all detectors to the edge computers, given that transferring data from the edges to the system's final nodes increases latency. However, this is not always practical, as IoT edge computers often have limited hardware capabilities. Running a complex algorithm on an edge node may be infeasible or slow, so a more efficient strategy could involve executing it on a node with superior computational power, while simpler detectors operate on the edge computer. Conversely, it may sometimes be beneficial to execute the complex algorithm on the edge computer first, as even with longer runtime, it may enhance overall detection performance. Our approach involves initially optimizing each node individually and then extending the optimization across the remaining nodes within the architecture.

\subsubsection{Single node}
Consider a collection of detectors $D$ that is available for selection at a given node, with each detector $d \in D$ possessing a performance score $p_d$ and execution time $t_d$. The aim is to enhance the node's overall performance, ensuring that the total execution time of the detectors does not exceed $T_{max}$. This constraint is imposed by the developer, as delays in triggering an alert would be deemed excessive. The problem can be redefined as follows:
\begin{equation}
    p_{node} = \max \sum_{i=1}^{|D|} x_i p_i
\end{equation}
subject to: 
\begin{equation}
     f(D' := \{d_i | \forall d_i \text{ where } x_i = 1\}) + t \leq T_{max}
\end{equation}
where $x_i \in \{0, 1\}$ and $f(\cdot)$ is a function that estimates the duration needed to operate the detectors at the node, with $t$ denoting the delay in receiving data from the prior nodes. This challenge is also known as the Knapsack problem \cite{kellerer_multidimensional_2004}, classified as NP-Hard. We selected F1 as our performance metric. However, other metrics such as Precision or Recall are also applicable. Consider the following simple example: we have three detectors, d1 (F1 = 0.6), d2 (F1 = 0.98), and d3 (F1 = 0.7). If the combined execution time of d1 and d3, when run together, is shorter than the execution time of d2 alone, then it is preferable to deploy d1 and d3. Conversely, if d1 and d3 cannot be parallelized to be shorter, we should instead deploy only d2.

In estimating the time required for the detectors to execute in the node, one might assume sequential execution. Nonetheless, this approach lacks optimization, necessitating some level of parallelization. Moreover, they should adhere to CPU ($cpu_{max}$) and RAM ($ram_{max}$) usage constraints, as other processes also need to operate on the node, or it may be a mobile node where excessive resource consumption could diminish battery life. Consequently, we compute $f(\cdot)$ as:

\begin{equation}
    %f(D') = \min \sum_{i=1}^{|D'|}\sum_{j=1}^{T_{max}} s_{i, j} 
    f(D') = \min \max_j s_{i, j}, \forall (i,j) \in (|D'|, \mathbb{N}), j < T_{max} 
\end{equation}
subject to: 
\begin{equation} 
    \sum_{i=1}^{|D'|} s_{i, j}d_{i}^{cpu} \leq cpu_{max}, \forall j \in T_{max}
\end{equation}
and:

\begin{equation} 
    \sum_{i=1}^{|D'|} s_{i, j} d_{i}^{ram} \leq ram_{max}, \forall j \in T_{max}
\end{equation}

where $s_{i, j} = \{0, 1\}$, we presume that a detector will continue operating once initiated until it concludes. To simplify, for a detector $d_i$, we determine $d_i^{cpu}$ and $d_i^{ram}$ by taking their peak values throughout the entire process. This problem is a variant of the project scheduling problem \cite{liu_project_2009}.

\subsubsection{All nodes}
After understanding the optimization of an individual node, the next step is to optimize the whole system. Algorithm \ref{alg:optimization} illustrates the system optimization process. Initially, we perform a topological sort on the tree to generate a list of layers (line \ref{alg:opt:top}), ensuring that each child layer is processed before its parent. During the loop, we apply distinct optimization strategies to each layer based on its type:
\begin{itemize}
    \item \textbf{Leaf layer}: All nodes in a leaf layer are composed of the same hardware and function independently. Therefore, by optimizing a single node, we can determine the detectors employed in the layer (line \ref{alg:opt:child}). To ensure compatibility with the data produced in the leaf, we initially screen out any detectors that are not suitable for use in the subsequent layers (line \ref{alg:opt:filt}).
    \item \textbf{Standard layer}: For a standard layer, we begin by adding the detectors that remain after excluding those already utilized by their child layers (line \ref{alg:opt:rem}). Subsequently, we optimize each node in the layer (line \ref{alg:opt:all}), ensuring that each node uses distinct detectors.
\end{itemize}
The procedure should be repeated until all layers have been processed. It is important to note that the algorithm does not ensure that every detector will be used in every scenario. However, it cannot be ensured that every detector is capable of being utilized. The \textit{opt} function employed in Algorithm \ref{alg:optimization} refers to optimization at the node level. Additionally, there may be instances where the same detector operates on both the child and parent layers. This can happen when a detector runs in child A but not in child B, even though it is required for both. We could incorporate checks to ensure that the parent detector does not reprocess data originating from child A. However, for simplicity's sake, this check was not implemented in this paper.
\begin{algorithm}
\caption{Node optimization}
\begin{algorithmic}
\STATE $layers$ $\gets$ \texttt{topological\_sort}($tree$) \label{alg:opt:top}
\STATE $layer\_d \gets \{\}$
\FOR{$layer$ \textbf{in} $layers$}
    \IF{\texttt{is\_leaf\_layer($layer$)}}
        \STATE $D_{cand} \gets $ \texttt{filter\_by\_type($D'$)} \label{alg:opt:filt}
        \STATE $D' \gets$ \texttt{opt}($D_{cand}$, $node := layers[0]$) \label{alg:opt:child}
    \ELSE
        \STATE $D_{cand} \gets \emptyset$
        \FOR{$child\_layer$ \textbf{in} \texttt{get\_children}($layer$)}
            \STATE $D_{cand} \gets D_{cand} \cup layer\_d[child\_layer]$ \label{alg:opt:rem}
        \ENDFOR
        \STATE $D' \gets \emptyset$, $D_{temp} \gets D_{cand}$
        \FOR{$node$ \textbf{in} $nodes$}
            \STATE $D_{temp} \gets $\texttt{opt}($D_{temp}$, $node$) \label{alg:opt:all}
            \STATE $D' \gets D'\cup D_{temp}$ 
        \ENDFOR
    \ENDIF
    \STATE $layer\_d$[$layer$] $\gets D_{cand} - D'$
\ENDFOR
\end{algorithmic}
\label{alg:optimization}
\end{algorithm}

\section{UGV dataset}
\label{sec:robot_dog}

We designed a fictional scenario to generate the dataset for our unmanned ground vehicle (UGV) unit. In this scenario, the unit patrols around a critical infrastructure site in challenging terrain, using an onboard camera. The attackers’ objective is to gain virtual and physical access to the vehicle so it can later be used for lateral movement attacks once it re-enters the infrastructure or other sabotage activities. As we will describe later, the network is composed of multiple interconnected nodes whose overall distribution may change under normal operating conditions. The UGV may leave the primary network, become isolated, or change to a different networks (inside and outside the infrastructure), thus altering the overall distribution.

\subsection{Network topology}

\begin{figure}[h]
    \centering
    \includegraphics[width=7.cm]{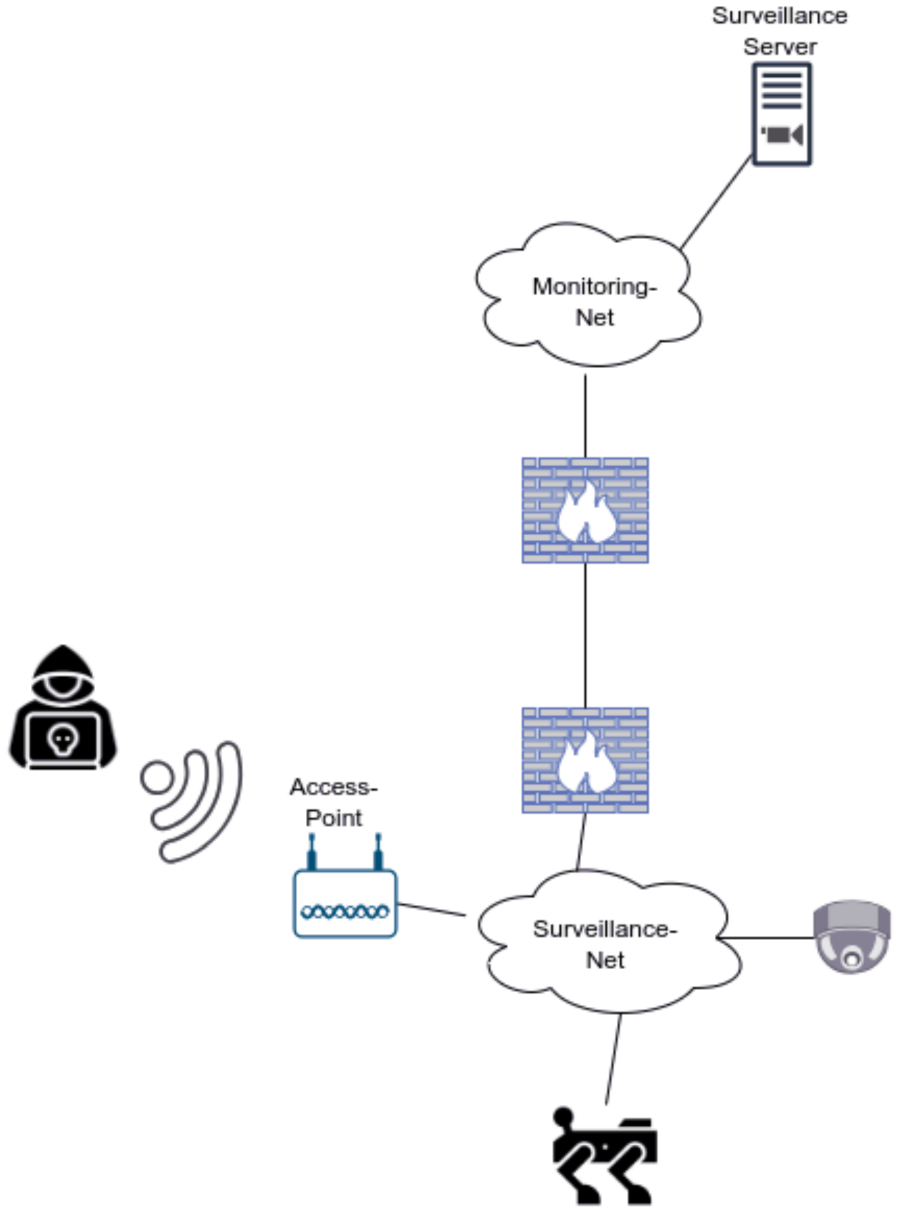}
    \caption{Network topology of the UGV use case.}
    \label{fig:robot_dog}
\end{figure}

Figure \ref{fig:robot_dog} shows the network topology employed to generate the dataset, which is based on multiple interconnected nodes. A robot dog unit and a camera monitor a difficult-to-reach area outside a critical infrastructure. Both devices are connected via Wi-Fi, and within a highly secured internal network inside the infrastructure, a surveillance server operates to collect the gathered information.

\subsection{Attack playbook}
The attackers use stolen credentials to first gain access to the Wi-Fi access point and, once inside, scans the network to locate and identify connected devices. Then they launch DoS attacks to blind both the surveillance camera and the robot dog unit to keep the physical access undetected. Once this initial phase is finished, a second attacker makes physical contact with the robot to install new hardware intended for sabotage operations, while the first attackers exploit a vulnerability in the robot dog that enables them to seize control of the unit. The attack sequence following the IDs of the ATT\&CK\footnote{Read 14-01-2026 https://attack.mitre.org/tactics/enterprise/} technique is: T1669,T1078, T1595, T1498, T1059.006,T1068, T1059, T1033, T1003.008, T1120, T1124. Full dataset can be found in \cite{hotwagner_2026_18482763}.

\subsection{Generation setup}
To build the dataset, we used a quadruped robotic dog platform (Figure \ref{fig:cluster}). The video camera is connected to a video server running ZoneMinder\footnote{https://zoneminder.readthedocs.io/en/latest/}. The attack sequence was carried out using the attack automation framework AttackMate \cite{landauer2026attackmate}.

\subsection{Dataset preprocessing}
To build the datasets, we use the logs from the robot dog unit. In this study, we applied Drain \cite{he_drain_2017} to preprocess these logs before executing the various log anomaly detection methods described in Section \ref{sec:experimental_setup}. To capture its typical behavior, we allowed the unit to operate for an hour without any interaction.

\section{Experimental Setup}
\label{sec:experimental_setup}

In this section, we present the experimental setup along with all the different components used.

\subsection{Datasets}

The experiments used three dataset categories representing different distributed use cases and data types, each necessitating a distinct type of IDS. All datasets were evaluated within the same CIDS framework, varying only the IDS used.
\begin{itemize}
    \item \textit{Unmanned Aerial Vehicles (UAVs)}: The data sets referenced \cite{hassler_cyber-physical_2024} focus on capturing air-to-ground communication traffic between a UAV and a Ground Control Station (GCS).
    \item \textit{Ground Infrastructure}: We utilized the AIT-LDv2 dataset \cite{landauer_maintainable_2023} to simulate attacks targeting multiple servers across different organizations. In this study, we specifically relied on the subset corresponding to the "fox" organization described in \cite{landauer_maintainable_2023}.
    \item \textit{Unmanned Ground Vehicles (UGVs)}: This dataset was developed for this study (Section \ref{sec:robot_dog}) \cite{hotwagner_2026_18482763}. 
\end{itemize}
The sizes of the different datasets are presented in Table \ref{tab:dataset_size}. We refer to Negatives as input instances such as logs produce by the system that do not reflect an attack, and Positives as those that do.
\begin{table}[t]
    \centering
    \caption{Size datasets}
    \label{tab:dataset_size}
    \begin{tabular}{l|c|c}
        \textbf{Dataset} & \textbf{Negatives} & \textbf{Positives} \\
        \hline
        \hline
        \textbf{Drone DoS} & 9425 & 11671 \\
        \textbf{Drone FDI} & 9425 & 3473 \\
        \textbf{Drone Replay} & 9425 & 12006 \\
        \textbf{AIT-LDv2 fox} & 2670 & 118 \\
        \textbf{Robotdog} & 18016 & 11413 \\
    \end{tabular}
\end{table}

\subsection{Optimization methods}

We employ metaheuristics to address the optimization problems because they are computationally efficient in discrete, low-dimensional search spaces. Integer Linear Programming and related techniques were not used, as they are not straightforward to implement in this context because the time constraint is computed dynamically and depends on the specific combination of IDS.

\begin{itemize}
    \item \textit{Local search (LS)}: Arguably the simplest metaheuristic, it starts from a given point in the search space and explores its neighborhood to find the best possible neighbor \cite{rego_local_2007}.
    \item \textit{Tabu search (TS)}: An enhanced variant based on LS \cite{laguna_tabu_2025} is capable of escaping certain local minima in which standard LS would become trapped.
    \item \textit{Ant Colony Optimization (ACO)}: We use the ACO algorithm to solve our optimization problem following the procedure 0-1 integer programming described in \cite{zhao_new_2006}.
\end{itemize}

To enable a comparison of the different methods against the optimal result, we additionally implemented a brute-force approach, using a dynamic programming breadth-first search strategy. All metaheuristics were configured with a patience of 2 iterations and were executed for up to 10 iterations in total.

\subsection{Intrusion Detection Methods}

For the real-world data experiments, we employed several types of IDS to achieve a more realistic setting. Each dataset is associated with its own set of intrusion detection methods. We chose methods known to be computationally lightweight so they can be efficiently executed on edge devices.
\subsubsection{UAV IDS}
We chose some supervised approaches presented in \cite{hassler_cyber-physical_2024}, where each method is trained to identify a particular attack using the given labels:
\begin{itemize}
    \item \textit{1D CNN}: A basic one-dimensional convolutional neural network (CNN) was implemented to identify Denial of Service (DoS) attacks. The hyperparameter settings are listed in Table \ref{tab:cnn}.
    \item \textit{MLP}: A fully connected neural network was employed to identify false data injection (FDI) attacks. Table \ref{tab:mlp} presents the hyperparameters used.
    \item \textit{Decision tree (DT)}: A decision tree with a maximum depth of 3 was applied to identify replay attacks.
\end{itemize} 

\begin{table}[t]
    \centering
    \caption{CNN architecture}
    \label{tab:cnn}
    \begin{tabular}{l|c}
        \textbf{Hyperparameters} & \textbf{Values} \\
        \hline
        \hline
        \textbf{Layer 1: CNN} & 34x32x3 relu \\
        \textbf{Layer 2: Max pool} & 2 \\
        \textbf{Layer3: CNN} & 32x64x3 relu \\
        \textbf{Layer 4: Max pool} & 2 \\
        \textbf{Layer 5: Aptative avg. pool} & 1 \\
        \textbf{Layer 6: Flatten} &  \\
        \textbf{Layer 7: Linear} & 64x128 relu \\
        \textbf{Layer 8: Dropout} & 0.5 \\
        \textbf{Layer 9: Linear} & 128x1 \\
        \hline
        \textbf{Loss} & Binary Cross-Entropy\\
        \textbf{Optimizer} & Adam lr=0.001 \\
        \textbf{Epochs / Batch size} & 20, 16 \\
    \end{tabular}
\end{table}

\begin{table}[t]
    \centering
    \caption{MLP architecture}
    \label{tab:mlp}
    \begin{tabular}{l|c}
        \textbf{Hyperparameters} & \textbf{Values} \\
        \hline
        \hline
        \textbf{Layer 1: Linear} & 34x32 relu \\
        \textbf{Layer 2: Linear} & 32x64 relu \\
        \textbf{Layer 3: Linear} & 64x1 \\
        \hline
        \textbf{Loss} & Binary Cross-Entropy\\
        \textbf{Optimizer} & Adam lr=0.001 \\
        \textbf{Epochs / Batch size} & 15, 10 \\
    \end{tabular}
\end{table}

\subsubsection{Ground Infrastructure IDS}

We implement our anomaly detection methods inspired by \cite{beck_semi-supervised_2024}, where they had previously achieved good results.
\begin{itemize}
    \item \textit{New values (NV)}: Monitor the logs for previously unseen variables to identify abnormal behavior.
    \item \textit{Entropy values (EV)}: Analyze how often these new variables appear to detect abnormal behavior.
\end{itemize}

\subsubsection{UGV IDS}

For our proposed dataset, we adopt an approach similar to that of \cite{landauer_critical_2024} to develop our anomaly detection techniques.
\begin{itemize}
    \item \textit{Edit distance (ED)}: this method relies on the Levenshtein distance as a metric to identify anomalies within the log sequence.
    \item \textit{Event cluster (EC)}: this method clusters log sequences based on similarity of event counts to detect abnormal behavior.
\end{itemize}

\subsection{Hardware setup}

To effectively simulate an IoT environment, five Raspberry Pi 5 devices were utilized. We chose this hardware to demonstrate that our approach could indeed operate on edge computers in a reasonable amount of time. The Raspberry Pi units utilized featured a 64-bit Arm Cortex-A76 processor, 8 GB of RAM, and were running Ubuntu 25.04-25.10 operating system. To ensure reproducibility, a Docker Compose file was included so the various nodes can be run on a local machine.

\begin{figure}[h]
    \centering
    \includegraphics[width=9cm]{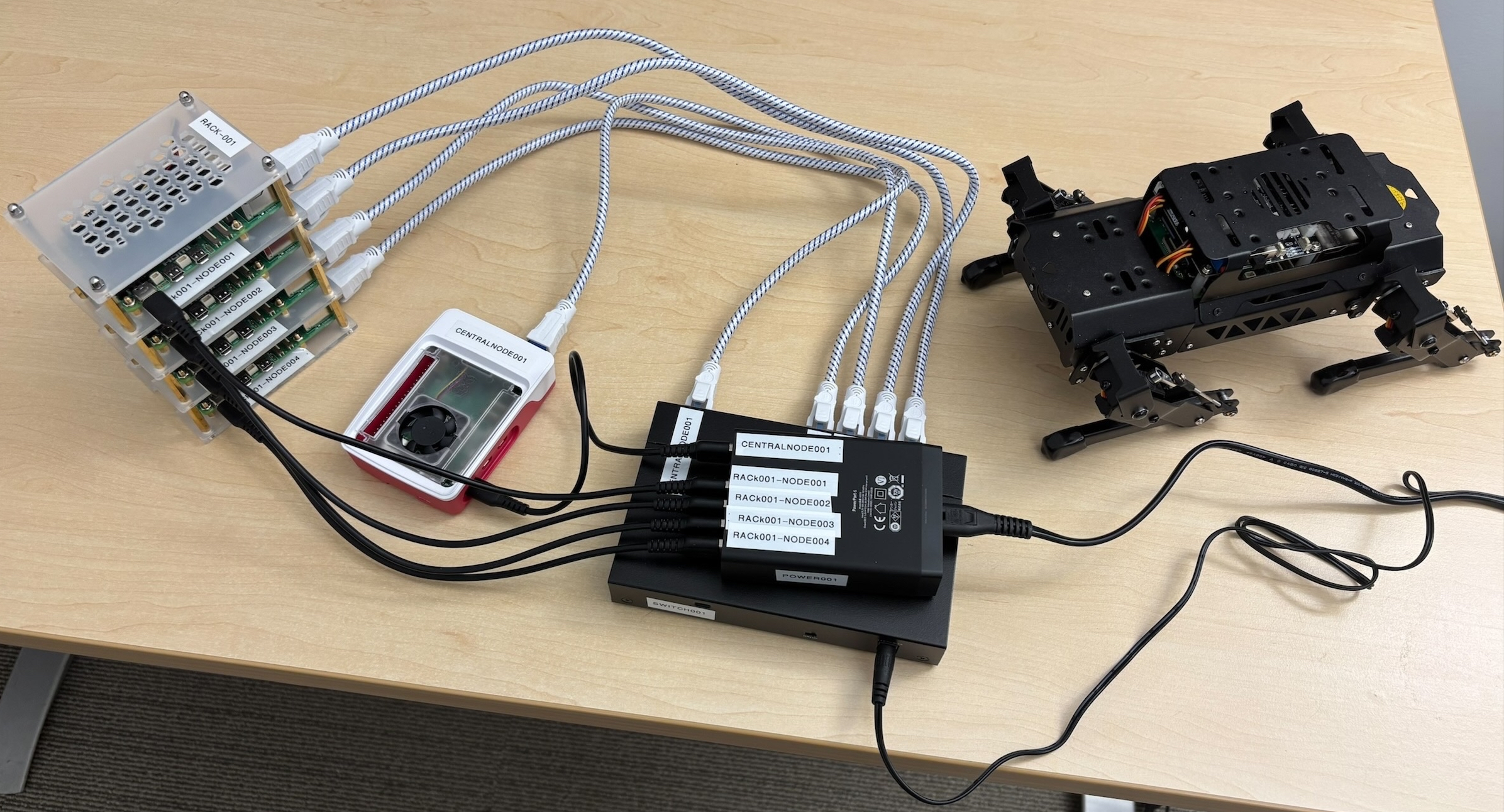}
    \caption{Hardware setup used in the paper. On the left is the edge computing cluster, and on the right is the robot dog unit.}
    \label{fig:cluster}
\end{figure}

Figure \ref{fig:cluster} illustrates the hardware configuration. The Raspberry Pi devices were arranged in an edge computing cluster, with all units connected to a switch that formed an internal network and to a power bank. The robot dog unit was used solely for data collection for the dataset and was not employed to run the different experiments.

\section{Results}
\label{sec:results}

As described in Section \ref{sec:experimental_setup}, the experiments are organized to obtain results for the different RQs and to showcase the effectiveness of the proposed CIDS framework. 

\subsection{Experiment with synthetic generated data}

We implemented a script designed to generate a random tree structure, adjusting variables such as the number of nodes, IDS, and data types. Within this structure, each node is outfitted with emulated hardware, with specifications like CPU and RAM being randomly determined. The value ranges for the various layer features are listed in Table \ref{tab:rand_node}, and those for the IDS are presented in Table \ref{tab:rand_ids}. Leaf nodes and IDS can each assume one of 6 randomly generated possible data types. The probability that a node is a standard node is computed using Equation \ref{eq:gen}. We employ this equation to produce a sparse tree, in which the number of branches decreases as the tree becomes deeper. The main motivation is to enable the construction of deep trees without causing an exponential growth in the total number of nodes. The $\alpha$ parameter lets us control how many nodes the equation will produce in the tree.
\begin{equation}
    \label{eq:gen}
        p(x) = 
    \begin{cases}
        e^{-\alpha x} & \text{if } x > x_{max}, \\
        0,     & \text{otherwise}
    \end{cases},
\end{equation}

where $x$ denotes the current number of layers, $x_{max}$ is the maximum value. Based on our empirical studies, we decided to set $x_{max} = 30$ and $\alpha = 0.2$.

\begin{table}[t]
    \centering
    \caption{Layer range values features}
    \label{tab:rand_node}
    \begin{tabular}{l|c}
        \textbf{Features} & \textbf{Range of values} \\
        \hline
        \hline
        \textbf{N. children} & [1 - 3] \textit{*(only standard nodes)} \\
        \textbf{N. nodes} & [1 - 4] \\
        \textbf{Computer} & \{Raspberry Pi 4, Orange pi, Banana pi\} \\
        \textbf{Max CPU} & [1 - 100] \% \\
        \textbf{Max Memory} & [512 - 8191] Bytes \\
        \textbf{Max Time} & [10 - 21] s \\ 
    \end{tabular}
\end{table}

\begin{table}[t]
    \centering
    \caption{IDS range value features}
    \label{tab:rand_ids}
    \begin{tabular}{l|c}
        \textbf{Features} & \textbf{Range of values} \\
        \hline
        \hline
        \textbf{F1 score} & [0.1 - 1.0] \\
        \textbf{CPU usage} & [1 - 31] \% \\
        \textbf{Memory usage} & [64-4096] Bytes \\
        \textbf{Time} & [1 - 10] s \\ 
    \end{tabular}
\end{table}

To accurately assess performance, we compared how the different metaheuristics performed against the brute-force method at each node, as shown in the following equation:
\begin{equation}
    M(f_m(N_T, \bigcup_{c \in N_T} c.IDS), f^*(N_T, \bigcup_{c \in N_T} c.IDS)) \rightarrow [0, 1],
    \label{eq:metric}
\end{equation}

where $M$ represents the metric used for comparison, specifically the F1 score, while $N_T$ denotes the current node, and $c.IDS$ signifies the collection of IDS that were transmitted to $N_T$ from the child nodes. We generated a total of 20 distinct architectures, each with a varying number of IDS, to evaluate overall performance. Since our optimization equation consists of two components, it necessitates two distinct optimization processes. We applied the metaheuristics and brute force to the knapsack problem. Running brute-force for scheduling was also very time-consuming, and our preliminary experiments showed that using a heuristic for scheduling led to only a negligible performance loss within the brute-force search. Therefore, at this stage, all scheduling was performed using Local Search to maintain consistency between runs.

\begin{figure}[h]
    \centering
    \includegraphics[width=9cm]{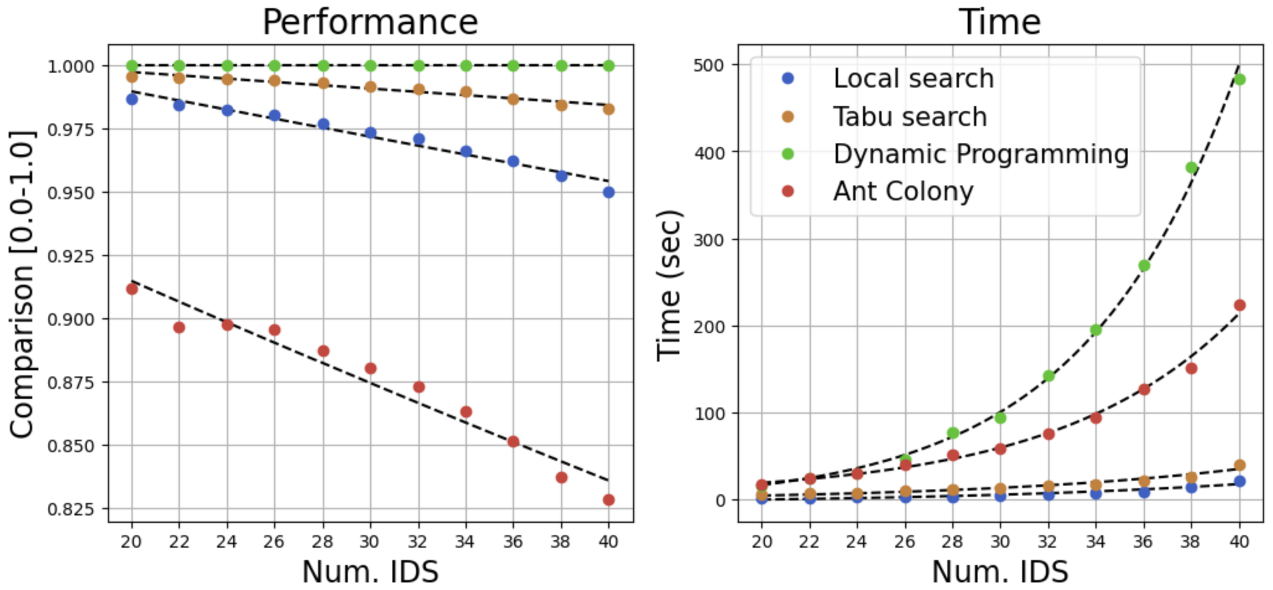}
    \caption{Comparison of overall performance and time between different amount of IDS for the Node Optimization.}
    \label{fig:performace_vs_time}
\end{figure}

\begin{figure}[h]
    \centering
    \includegraphics[width=9cm]{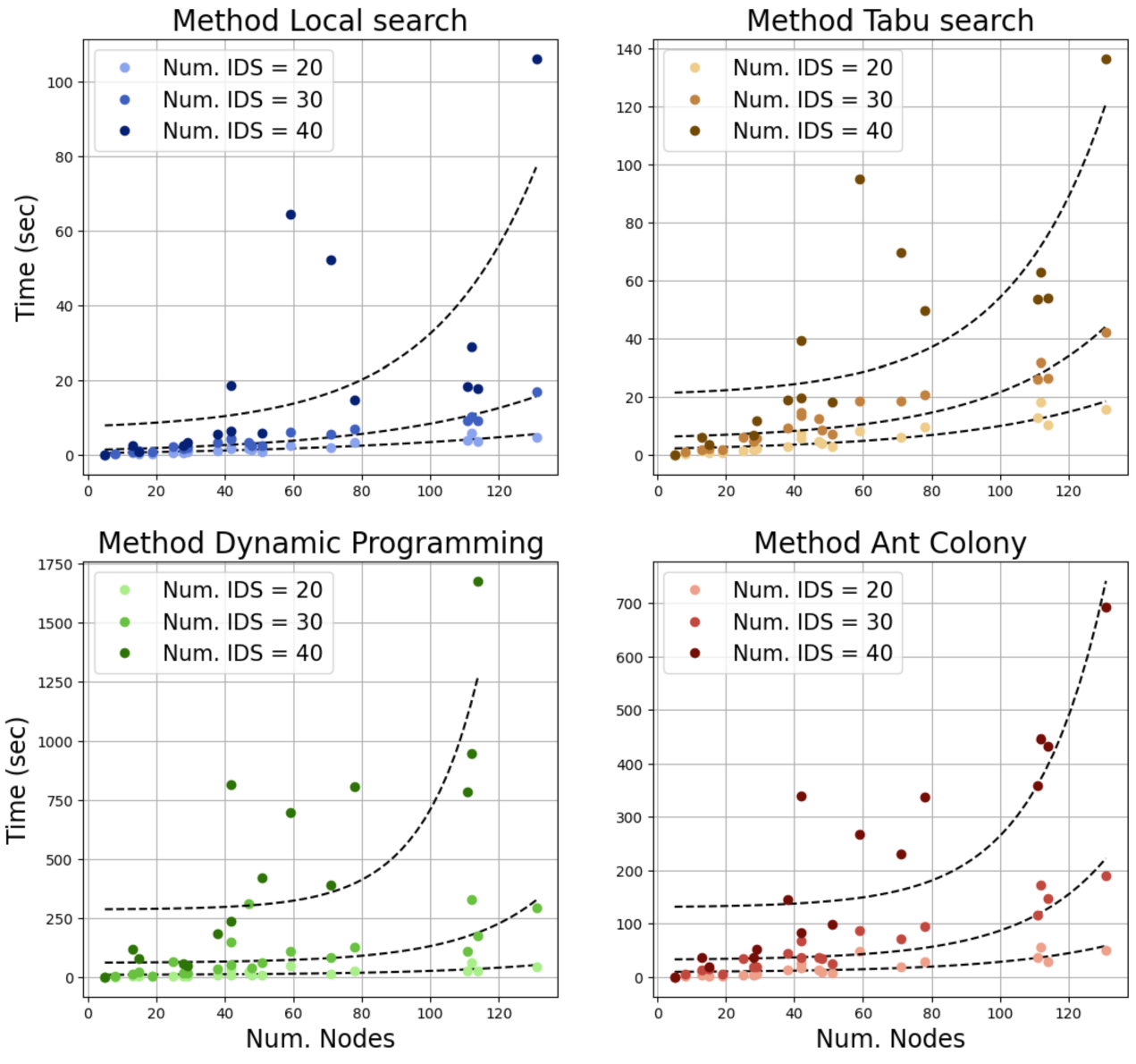}
    \caption{Comparison of overall performance and time between different amount of nodes for the Node Optimization.}
    \label{fig:performace_vs_time_all}
\end{figure}

Two figures are used to analyze the solvers performance. Figure \ref{fig:performace_vs_time} shows the trade-off between computational time and solution performance (per Equation \ref{eq:metric}). Each point is the average result of 20 architectures using different solver methods, with 20–40 IDS chosen as an appropriate range. The results show that Tabu search achieves the highest solution performance compared to the other metaheuristics, closely approaching the optimal solutions obtained via brute-force dynamic programming. Notably, both Tabu search and Local search show minimal increases  in computation time as the number of nodes grows, demonstrating scalability. Figure \ref{fig:performace_vs_time_all} presents how various metaheuristics scale with the number of IDS and architectural nodes, each point being the average of three timing measurements.

\subsection{Experiment with real data}

\begin{figure}[h]
    \centering
    \includegraphics[width=7.5cm]{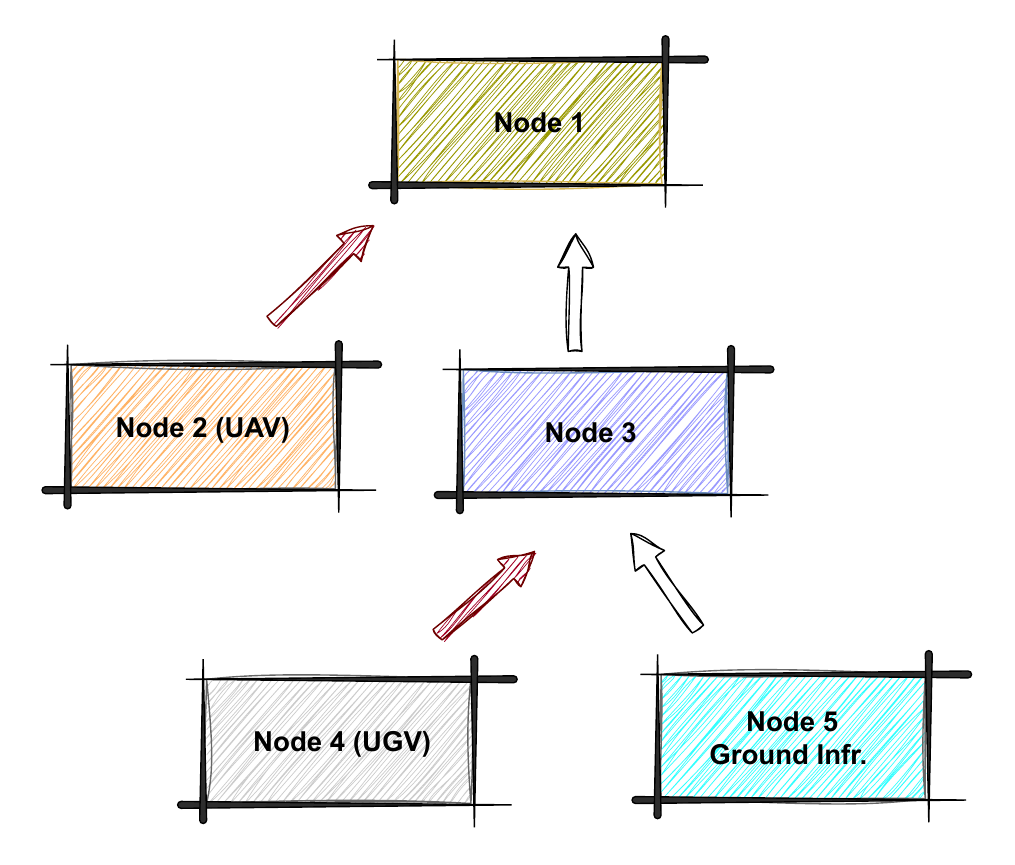}
    \caption{Test scenario}
    \label{fig:real_case}
\end{figure}

We construct a test scenario in which our CIDS framework is deployed on an architecture consisting of five edge machines (Section \ref{sec:experimental_setup}). This setup includes three leaf nodes (each associated with a distinct dataset) and two standard nodes. The aim of this experiment is to illustrate how the framework functions in a realistic environment and to assess its performance. Figure \ref{fig:real_case} depicts the distribution of the nodes as a simplified version of Figure \ref{fig:example_cids} where each layer is composed of a single node. The experiment is organized into different phases:
\begin{itemize}
    \item \textbf{(Phase 1) Training}: The first phase focuses on training the different IDS instances. Because, in this setup, all nodes are part of the same organization, no federated training is used. Instead, each leaf node independently trains its own IDS using its locally available data. Once training is finished, the resulting model statistics are propagated to the upper-level nodes.
    \item \textbf{(Phase 2) First optimization}: The first optimization step is performed with all nodes connected. For this, we adopt Tabu search as the primary metaheuristic technique.
    \item \textbf{(Phase 3) Second optimization}: In the second optimization phase, Node 2 and Node 4 are isolated from the remaining nodes, as indicated by the red arrows in Figure \ref{fig:real_case}. Under these conditions, the nodes detect the loss of connectivity and adjust their behavior accordingly.
    \item \textbf{(Phase 4) Analysis}: After collecting the results, an analysis is performed to assess the overall performance of the architecture and the hardware resources allocated to each node.
\end{itemize}

Table \ref{tab:specifications} summarizes the chosen specifications for each node in the experiment, including maximum CPU percentage, RAM size, and execution time allowed for running all detectors on a single input. These settings were selected to demonstrate the CIDS framework’s behavior, but alternative specifications could also be used. 

\begin{table}[t]
    \centering
    \caption{Node specifications, increase done in phase 2 was added in parenthesis.}
    \label{tab:specifications}
    \begin{tabular}{l|c|c|c}
        \textbf{Node} & \textbf{Max time (ms)} & \textbf{Max CPU (\%)} & \textbf{Max RAM (MB)} \\
        \hline
        \hline
        \textbf{Node 1} & 100 & 100 & 1000 \\
        \textbf{Node 2} & 200 & 50 & 3  (+60) \\
        \textbf{Node 3} & 200 & 50 & 4\\
        \textbf{Node 4} & 200 (+300)& 50 (+10) & 50 \\
        \textbf{Node 5} & 200 & 50 & 3 \\
    \end{tabular}
\end{table}

\subsubsection{Training}

Training for the anomaly detection methods uses 10\% of the data, while 80\% is allocated to the supervised methods. We run each method three times and report the mean and standard deviation. For performance, we use F1, Precision, and Recall metrics. We also measure the time required to process a single input in milliseconds, as well as the maximum CPU usage and RAM consumption (in MB). Since each dataset has a different size, we also report the Wilson score intervals for Recall and Precision with a 95\% confidence level. Table \ref{tab:specifications} and Table \ref{tab:hardware_spec} present the performance of the different methods.

\begin{table}[t]
    \centering
    \caption{Intrusion detection methods performance benchmark}
    \label{tab:specifications}
    \begin{tabular}{l|c|c|c|c|c}
        \textbf{Name} & \textbf{F1} & \textbf{Recall} & \textbf{Recall c.} &\textbf{Prec.} & \textbf{Prec. c.}\\
        \hline
        \hline
         \textbf{CNN} & 0.94±0.00 & 0.89±0.00 & 0.88-0.91 & 0.98±0.00& 0.97-0.99 \\
         \textbf{MLP} & 0.43±0.02 & 1.00±0.00 & 1.00-1.00 & 0.27±0.00 & 0.26-0.29 \\
         \textbf{DT} & 0.89±0.00 & 0.81±0.01 & 0.79-0.82 & 0.98±0.01 & 0.98-0.99\\
        \hline
         \textbf{NV} & 0.94±0.04 & 0.95±0.04 & 0.90-0.98 & 0.93±0.04 & 0.87-0.96\\
         \textbf{EV} & 0.93±0.01 & 0.88±0.02 & 0.81-0.93 & 0.97±0.02 & 0.92-0.99\\
        \hline
         \textbf{ED} & 0.66±0.01  & 0.61±0.01 & 0.60-0.62 & 0.71±0.01 & 0.70-0.72\\
         \textbf{EC} & 0.61±0.00 & 0.68±0.01 & 0.67-0.69 & 0.55±0.00 & 0.54-0.56\\
    \end{tabular}
\end{table}
% Missing only Precision std!!!
\begin{table}[t]
    \centering
    \caption{Intrusion detection methods hardware benchmark}
    \label{tab:hardware_spec}
    \begin{tabular}{l|c|c|c}
        \textbf{Name} & \textbf{Time (ms)} & \textbf{Max CPU (\%)} & \textbf{Max RAM (MB)} \\
        \hline
        \hline
         \textbf{CNN} & 3.14 & 8.07 \% & 57.73 \\
         \textbf{MLP} & 2.56 & 7.65 \% & 0.58 \\
         \textbf{DT} & 1.79 & 8.03 \% &  1.48 \\
        \hline
         \textbf{NV} & 6.78 & 2.73 \% & 0.00 \\
         \textbf{EV} & 8.59 & 2.65 \% & 0.00 \\
        \hline
         \textbf{ED} & 8.08 & 8.91 \% & 33.15 \\
         \textbf{EC} & 27.20 & 9.17 \% & 78.38 \\
    \end{tabular}
\end{table}

\subsubsection{Optimizations}

The runtime of each phase is presented in Table \ref{tab:times}. To achieve results with higher statistical significance, the optimization phase is run three times. Since the training process in the previous phase was also performed three times, we divided the reported time by three. To emulate a realistic scenario in which the changes occurring in phase 2 have an impact on the system, we configure Node 1 and Node 4 to increase their resource usage within the framework when they become disconnected from the rest of the architecture. The core rationale is that, in such a situation, battery saving should become a lower priority, since these nodes no longer benefit from the support of the overall architecture and must instead execute more detectors locally. The increase done can be found in Table \ref{tab:specifications}. Table \ref{tab:methods} presents the detectors chosen for each node.

\begin{table}[t]
    \centering
    \caption{Time for the different phases}
    \label{tab:times}
    \begin{tabular}{l|c|c|c}
        \textbf{Node} & \textbf{Train (s)} & \textbf{First optim. (s)} & \textbf{Second optim. (s)} \\
        \hline
        \hline
        \textbf{Node 1} & - & 0.07±0.00 & 0.03±0.00 \\
        \textbf{Node 2} & 316.94 & 0.04±0.01 & 0.06±0.01 \\
        \textbf{Node 3} & - & 0.03±0.01 & 0.02±0.00 \\
        \textbf{Node 4} & 1021.65 & 0.03±0.01 & 0.03±0.01 \\
        \textbf{Node 5} & 53.44 & 0.05±0.01 & 0.06±0.00 \\
    \end{tabular}
\end{table}

\begin{table}[h!]
    \centering
    \caption{Methods selected in phase 2 and phase 3}
    \label{tab:methods}
    \begin{tabular}{l|c|c}
        \textbf{Node} & \textbf{Phase 1 methods} & \textbf{Phase 2 methods} \\
        \hline
        \hline
        \textbf{Node 1} &  EC, CNN & - \\
        \textbf{Node 2} & MLP, DT & MLP, DT, CNN \\
        \textbf{Node 3} & - & - \\
        \textbf{Node 4} & ED & ED \\
        \textbf{Node 5} & NV, EV & NV, EV \\
    \end{tabular}
\end{table}

\subsubsection{Analysis}
Table \ref{tab:methods} summarizes the results from the first and second phases of the experiment, from which the following main observations can be made:
\begin{itemize}
    \item \textit{Node 3}: In both phases, no detectors are deployed on Node 3, indicating that the resources allocated to this node are insufficient.
    \item \textit{Phase 1}: The first phase indicates that the system as a whole provides enough resources to deploy all detectors.
    \item \textit{Phase 2}: The second phase shows that the countermeasures available on Node 2 are sufficient to deploy all detectors locally. This does not hold for Node 4, where, even after increasing its available resources, it still deploys the same number of detectors locally.
\end{itemize}
Based on these results, we can draw conclusions and reallocate resources among the different nodes. In a real-world scenario, a risk assessment could be carried out to identify the most frequent or most critical situations and to examine how the CIDS framework would respond, allowing us to refine resource usage at each node. This analysis could also reveal that certain nodes are redundant and may not be necessary (for example, Node 3).
\section{Results Discussion}
\label{sec:discussion}

 \begin{tikzpicture}
  % Draw rectangle
  \draw (0,0) rectangle (8,4);

  % Put a paragraph inside the rectangle (automatically wrapped)
  \node[align=justify, text width=7.5cm] at (4,2.0) {
    \textbf{RQ1: How effectively does the CIDS framework operate across different edge device architectures?} The results indicate that the framework performs effectively across heterogeneous edge device architectures. Metaheuristics-based optimization enables can achieve performance comparable to a brute-force method in a relatively short time, even when using fairly large architectural configurations.
  };
\end{tikzpicture}

 As illustrated in Figures \ref{fig:performace_vs_time} and \ref{fig:performace_vs_time_all}, metaheuristic methods exhibit better time scalability, maintaining good solution quality with only a slight degradation in performance as the number of IDS increases. Among the methods evaluated, ACO underperformed. This is likely because ACO cannot exploit its full potential, as it is unable to assign values to individual detectors whose value depends on their combination with the others. This limits the information available to the ants and results in less optimal solutions compared to simpler metaheuristics. Furthermore, increasing the number of nodes and IDS instances leads to longer optimization times. \\

\begin{tikzpicture}
  % Draw rectangle
  \draw (0,0) rectangle (8,3.5);

  % Put a paragraph inside the rectangle (automatically wrapped)
  \node[align=justify, text width=7.5cm] at (4,1.8) {
    \textbf{RQ2: How well does the proposed framework perform when deployed in a realistic, multi-domain scenario?} Experiments using real-world data show that per-node optimization yield promising optimization times under hardware constraints. This enables rapid reconfiguration in response to environmental changes or security incidents.
  };
\end{tikzpicture}

The experiments on real-world data indicate that the optimization time for each node is minimal compared to the overall training time, allowing detectors to be quickly redeployed and reconfigured (Table \ref{tab:times}). Although network latency remains an external factor beyond control, this rapid optimization speed enhances resilience. For example, in the UGV attack scenario (Section \ref{sec:robot_dog}), the ground drone was isolated and became vulnerable to attacks. Our framework allows for such a case that the robot can quickly scale up and optimize the number of active IDS instances and operate its own standalone CIDS without depending on the complete architecture. This self-adaptive capability can improve survivability in contested environments.

\section{Threat to Validity}
\label{sec:threat}

The primary threat to validity is the limited number of use cases for which we evaluated our framework with real data. While a more extensive study lies beyond the scope of this paper, in future work we plan to devise additional scenarios and perform more rigorous testing. This will also cover key aspects that were not examined in RQ2, such as latency in the optimization step.

\section{Conclusion}
\label{sec:conclusion}

In this paper, we present a novel CIDS framework that can adapt to changes occurring in heterogeneous IoT environments. Motivated by the limitations of existing static CIDS architectures, our approach enables automatic reconfiguration of detector deployment across distributed nodes based on available resources, data modalities, and operational demands. The framework uses metaheuristic optimization to achieve near-optimal configurations in minimal time, making it suitable for deployment on resource-constrained edge devices.

To support realistic evaluation, we generated a new public UGV dataset capturing a multi-stage cyberattack. Through multiple experiments, we demonstrated that our CIDS framework can adjust to changes and redeploy its internal IDS within a short time. Thanks to its robust design, it tolerates network failures such as those observed in our empirical experiments, enhancing resilience in challenging environments.

In future work, we plan to design more complex scenarios and datasets, and to add new capabilities such as optimized federated training, prioritization queues between detectors and the inclusion of additional considerations, including latency, a detailed threat model for potential system attacks, and a more explicitly defined security and trust layer among nodes. While in this initial version we focused primarily on optimizing the deployment of the IDS within the system, we also aim to improve alert management by developing alert aggregation techniques that can adapt to network changes and configurations.

\section*{Acknowledgment}
Funded by the European Union under the European Defence Fund (GA no. 101121403 - NEWSROOM and GA no. 101121414 - LATACC). Views and opinions expressed are however those of the author(s) only and do not necessarily reflect those of the European Union or the European Commission. Neither the European Union nor the granting authority can be held responsible for them. The author(s) used the Writefull Overleaf extension for grammar enhncement and vscode Copilot as a support tool for code implementation.

\bibliographystyle{ieeetr}
\bibliography{references}

@inproceedings{zhao_new_2006,
	title = {A new ant colony optimization for the knapsack problem},
	url = {https://ieeexplore.ieee.org/abstract/document/4127097},
	doi = {10.1109/CAIDCD.2006.329439},
	urldate = {2025-12-05},
	booktitle = {7th {Int.} {Conf.} on {Comp.}-{Aided} {Industrial} {Design} and {Conceptual} {Design}},
	author = {Zhao, Peiyi and et al.},
	month = nov,
	year = {2006},
	pages = {1--3},
}

@inproceedings{he_drain_2017,
	title = {Drain: {An} {Online} {Log} {Parsing} {Approach} with {Fixed} {Depth} {Tree}},
	shorttitle = {Drain},
	url = {https://ieeexplore.ieee.org/document/8029742},
	doi = {10.1109/ICWS.2017.13},
	urldate = {2025-12-03},
	booktitle = {{IEEE} {Int.} {Conf.} on {Web} {Services} ({ICWS})},
	author = {He, Pinjia and et al.},
	month = jun,
	year = {2017},
	pages = {33--40},
}

@article{carlo_cyber_2024,
	title = {Cyber attacks on critical infrastructures and satellite communications},
	volume = {46},
	issn = {1874-5482},
	url = {https://www.sciencedirect.com/science/article/pii/S1874548224000428},
	doi = {10.1016/j.ijcip.2024.100701},
	urldate = {2025-11-27},
	journal = {Int. Journal of Critical Infrastructure Protection},
	author = {Carlo, Antonio and Obergfaell, Kim},
	month = sep,
	year = {2024},
	pages = {100701},
}

@article{md_haris_uddin_sharif_literature_2022,
	title = {A literature review of financial losses statistics for cyber security and future trend},
	volume = {15},
	issn = {25819615},
	url = {https://wjarr.com/content/literature-review-financial-losses-statistics-cyber-security-and-future-trend},
	doi = {10.30574/wjarr.2022.15.1.0573},
	language = {en},
	number = {1},
	urldate = {2025-11-27},
	journal = {World Journal of Advanced Research and Reviews},
	author = {{Md Haris Uddin Sharif} and et al.},
	month = jul,
	year = {2022},
	pages = {138--156},
}

@article{landauer_critical_2024,
	title = {A {Critical} {Review} of {Common} {Log} {Data} {Sets} {Used} for {Evaluation} of {Sequence}-{Based} {Anomaly} {Detection} {Techniques}},
	volume = {1},
	url = {https://dl.acm.org/doi/10.1145/3660768},
	doi = {10.1145/3660768},
	number = {FSE},
	urldate = {2025-11-26},
	journal = {Reproduction package for article},
	author = {Landauer, Max and et al.},
	month = jul,
	year = {2024},
	pages = {61:1354--61:1375},
}

@incollection{laguna_tabu_2025,
	title = {Tabu {Search}},
	isbn = {978-3-032-00385-0},
	url = {https://link.springer.com/rwe/10.1007/978-3-032-00385-0_24},
	language = {en},
	urldate = {2025-11-26},
	booktitle = {Handbook of {Heuristics}},
	publisher = {Springer, Cham},
	author = {Laguna, Manuel},
	year = {2025},
	doi = {10.1007/978-3-032-00385-0_24},
	pages = {941--958},
}

@incollection{rego_local_2007,
	address = {Boston, MA},
	title = {Local {Search} and {Metaheuristics}},
	isbn = {978-0-306-48213-7},
	url = {https://doi.org/10.1007/0-306-48213-4_8},
	language = {en},
	urldate = {2025-11-26},
	booktitle = {The {Traveling} {Salesman} {Problem} and {Its} {Variations}},
	publisher = {Springer US},
	author = {Rego, César and et al.},
	editor = {Gutin, Gregory and Punnen, Abraham P.},
	year = {2007},
	doi = {10.1007/0-306-48213-4_8},
	pages = {309--368},
}

@article{landauer_maintainable_2023,
	title = {Maintainable {Log} {Datasets} for {Evaluation} of {Intrusion} {Detection} {Systems}},
	volume = {20},
	issn = {1941-0018},
	url = {https://ieeexplore.ieee.org/abstract/document/9866880},
	doi = {10.1109/TDSC.2022.3201582},
	number = {4},
	urldate = {2025-11-03},
	journal = {IEEE Transactions on Dependable and Secure Computing},
	author = {Landauer, Max and et al.},
	month = jul,
	year = {2023},
	pages = {3466--3482},
}

@article{hassler_cyber-physical_2024,
	title = {Cyber-{Physical} {Intrusion} {Detection} {System} for {Unmanned} {Aerial} {Vehicles}},
	volume = {25},
	issn = {1558-0016},
	url = {https://ieeexplore.ieee.org/abstract/document/10368002},
	doi = {10.1109/TITS.2023.3339728},
	number = {6},
	urldate = {2025-11-03},
	journal = {IEEE Transactions on Intelligent Transportation Systems},
	author = {Hassler, Samuel Chase and et al.},
	month = jun,
	year = {2024},
	pages = {6106--6117},
}

@inproceedings{wurzenberger_newsroom_2024,
	address = {New York, NY, USA},
	series = {{ARES} '24},
	title = {{NEWSROOM}: {Towards} {Automating} {Cyber} {Situational} {Awareness} {Processes} and {Tools} for {Cyber} {Defence}},
	isbn = {979-8-4007-1718-5},
	shorttitle = {{NEWSROOM}},
	url = {https://dl.acm.org/doi/10.1145/3664476.3670914},
	doi = {10.1145/3664476.3670914},
	urldate = {2025-10-31},
	booktitle = {Proce. of the 19th {Int.} {Conf.} on {Availability}, {Rel.} and {Sec.}},
	publisher = {Association for Comp. Machinery},
	author = {Wurzenberger, Markus and et al.},
	year = {2024},
	pages = {1--11},
}

@misc{noauthor_deeplog_nodate,
	title = {{DeepLog} {\textbar} {Proce.} of the 2017 {ACM} {SIGSAC} {Conf.} on {Comp.} and {Communications} {Security}},
	url = {https://dl.acm.org/doi/10.1145/3133956.3134015},
	urldate = {2025-10-21},
}

@article{roschke_advanced_2010,
	title = {An {Advanced} {IDS} {Management} {Architecture}},
	language = {en},
	author = {Roschke, Sebastian and et al.},
	year = {2010},
}

@book{shostack_threat_2014,
	title = {Threat {Modeling}: {Designing} for {Security}},
	isbn = {978-1-118-81005-7},
	shorttitle = {Threat {Modeling}},
	language = {en},
	publisher = {John Wiley \& Sons},
	author = {Shostack, Adam},
	month = feb,
	year = {2014},
	note = {Google-Books-ID: YiHcAgAAQBAJ},
}

@article{landauer_dealing_2022,
	title = {Dealing with {Security} {Alert} {Flooding}: {Using} {Machine} {Learning} for {Domain}-independent {Alert} {Aggregation}},
	volume = {25},
	issn = {2471-2566},
	shorttitle = {Dealing with {Security} {Alert} {Flooding}},
	url = {https://dl.acm.org/doi/10.1145/3510581},
	doi = {10.1145/3510581},
	number = {3},
	urldate = {2025-10-17},
	journal = {ACM Trans. Priv. Secur.},
	author = {Landauer, Max and et al.},
	month = apr,
	year = {2022},
	pages = {18:1--18:36},
}

@article{chiba_automatic_2022,
	title = {Automatic {Building} of a {Powerful} {IDS} for {The}
Cloud {Based} on {Deep} {Neural} {Network} by {Using} a
Novel {Combination} of {Simulated} {Annealing}
Algorithm and {Improved} {Self}-{Adaptive} {Genetic}
Algorithm},
	volume = {14},
	journal = {Int. Journal of Communication Networks and Information Security (IJCNIS)},
	author = {Chiba, Zouhair and et al.},
	year = {2022},
}

@inproceedings{beck_semi-supervised_2024,
	title = {Semi-supervised {Configuration} and {Optimization} of {Anomaly} {Detection} {Algorithms} on {Log} {Data}},
	url = {https://ieeexplore.ieee.org/abstract/document/10825202},
	doi = {10.1109/BigData62323.2024.10825202},
	urldate = {2025-10-17},
	booktitle = {{IEEE} {Int.} {Conf.} on {Big} {Data} ({BigData})},
	author = {Beck, Viktor and et al.},
	month = dec,
	year = {2024},
	note = {ISSN: 2573-2978},
	pages = {2575--2585},
}

@article{al-fuqaha_internet_2015,
	title = {Internet of {Things}: {A} {Survey} on {Enabling} {Technologies}, {Protocols}, and {Applications}},
	volume = {17},
	issn = {1553-877X},
	shorttitle = {Internet of {Things}},
	url = {https://ieeexplore.ieee.org/document/7123563},
	doi = {10.1109/COMST.2015.2444095},
	number = {4},
	urldate = {2025-10-17},
	journal = {IEEE Communications Surveys \& Tutorials},
	author = {Al-Fuqaha, Ala and et al.},
	year = {2015},
	pages = {2347--2376},
}

@article{stellios_survey_2018,
	title = {A {Survey} of {IoT}-{Enabled} {Cyberattacks}: {Assessing} {Attack} {Paths} to {Critical} {Infrastructures} and {Services}},
	volume = {20},
	issn = {1553-877X},
	shorttitle = {A {Survey} of {IoT}-{Enabled} {Cyberattacks}},
	url = {https://ieeexplore.ieee.org/document/8410404/},
	doi = {10.1109/COMST.2018.2855563},
	number = {4},
	urldate = {2025-10-17},
	journal = {IEEE Communications Surveys \& Tutorials},
	author = {Stellios, Ioannis and et al.},
	year = {2018},
	pages = {3453--3495},
}

@article{garcia_gomez_collaborative_2026,
	title = {Collaborative anomaly detection in log data: {Comparative} analysis and evaluation framework},
	volume = {175},
	issn = {0167-739X},
	shorttitle = {Collaborative anomaly detection in log data},
	url = {https://www.sciencedirect.com/science/article/pii/S0167739X2500384X},
	doi = {10.1016/j.future.2025.108090},
	urldate = {2025-10-17},
	journal = {Future Generation Comp. Systems},
	author = {García Gómez, André and et al.},
	month = feb,
	year = {2026},
	pages = {108090},
}

@inproceedings{alrashdi_ad-iot_2019,
	title = {{AD}-{IoT}: {Anomaly} {Detection} of {IoT} {Cyberattacks} in {Smart} {City} {Using} {Machine} {Learning}},
	shorttitle = {{AD}-{IoT}},
	url = {https://ieeexplore.ieee.org/abstract/document/8666450},
	doi = {10.1109/CCWC.2019.8666450},
	urldate = {2025-10-17},
	booktitle = {2019 {IEEE} 9th {Annual} {Computing} and {Communication} {Workshop} and {Conf.} ({CCWC})},
	author = {Alrashdi, Ibrahim and et al.},
	month = jan,
	year = {2019},
	pages = {0305--0310},
}

@article{andreolini_collaborative_2015,
	series = {Security and privacy information technologies and applications for wireless pervasive computing environments},
	title = {A collaborative framework for intrusion detection in mobile networks},
	volume = {321},
	issn = {0020-0255},
	url = {https://www.sciencedirect.com/science/article/pii/S0020025515001905},
	doi = {10.1016/j.ins.2015.03.025},
	urldate = {2025-10-17},
	journal = {Information Sciences},
	author = {Andreolini, Mauro and et al.},
	month = nov,
	year = {2015},
	pages = {179--192},
}

@article{aviv_russian-ukraine_2023,
	title = {Russian-{Ukraine} armed conflict: {Lessons} learned on the digital ecosystem},
	volume = {43},
	issn = {1874-5482},
	shorttitle = {Russian-{Ukraine} armed conflict},
	url = {https://www.sciencedirect.com/science/article/pii/S1874548223000501},
	doi = {10.1016/j.ijcip.2023.100637},
	urldate = {2025-10-17},
	journal = {Int. Journal of Critical Infrastructure Protection},
	author = {Aviv, Itzhak and Ferri, Uri},
	month = dec,
	year = {2023},
	pages = {100637},
}

@incollection{liu_project_2009,
	address = {Berlin, Heidelberg},
	title = {Project {Scheduling} {Problem}},
	isbn = {978-3-540-89484-1},
	url = {https://doi.org/10.1007/978-3-540-89484-1_9},
	language = {en},
	urldate = {2025-08-19},
	booktitle = {Theory and {Practice} of {Uncertain} {Programming}},
	publisher = {Springer},
	author = {Liu, Baoding},
	editor = {Liu, Baoding},
	year = {2009},
	doi = {10.1007/978-3-540-89484-1_9},
	pages = {139--146},
}

@incollection{kellerer_multidimensional_2004,
	address = {Berlin, Heidelberg},
	title = {Multidimensional {Knapsack} {Problems}},
	isbn = {978-3-540-24777-7},
	url = {https://doi.org/10.1007/978-3-540-24777-7_9},
	language = {en},
	urldate = {2025-08-19},
	booktitle = {Knapsack {Problems}},
	publisher = {Springer},
	author = {Kellerer, Hans and et al.},
	editor = {Kellerer, Hans and Pferschy, Ulrich and Pisinger, David},
	year = {2004},
	doi = {10.1007/978-3-540-24777-7_9},
	pages = {235--283},
}

@misc{mcmahan_communication-efficient_2023,
	title = {Communication-{Efficient} {Learning} of {Deep} {Networks} from {Decentralized} {Data}},
	url = {http://arxiv.org/abs/1602.05629},
	doi = {10.48550/arXiv.1602.05629},
	urldate = {2025-05-16},
	publisher = {arXiv},
	author = {McMahan, H. Brendan and et al.},
	month = jan,
	year = {2023},
	note = {arXiv:1602.05629 [cs]},
}

@inproceedings{nguyen_diot_2019,
	title = {D{ÏoT}: {A} {Federated} {Self}-learning {Anomaly} {Detection} {System} for {IoT}},
	shorttitle = {D{ÏoT}},
	url = {https://ieeexplore.ieee.org/document/8884802},
	doi = {10.1109/ICDCS.2019.00080},
	urldate = {2025-06-24},
	booktitle = {{IEEE} 39th {Int.} {Conf.} on {Distributed} {Comp.} {Sys.} ({ICDCS})},
	author = {Nguyen, Thien Duc and et al.},
	month = jul,
	year = {2019},
	note = {ISSN: 2575-8411},
	pages = {756--767},
}

@article{li_federated_2022,
	title = {Federated {Anomaly} {Detection} on {System} {Logs} for the {Internet} of {Things}: {A} {Customizable} and {Communication}-{Efficient} {Approach}},
	volume = {19},
	issn = {1932-4537},
	shorttitle = {Federated {Anomaly} {Detection} on {System} {Logs} for the {Internet} of {Things}},
	url = {https://ieeexplore.ieee.org/document/9716881},
	doi = {10.1109/TNSM.2022.3152620},
	number = {2},
	urldate = {2025-06-24},
	journal = {IEEE Transactions on Network and Service Management},
	author = {Li, Beibei and et al.},
	month = jun,
	year = {2022},
	pages = {1705--1716},
}

@inproceedings{kholidy_cids_2012,
	title = {{CIDS}: {A} {Framework} for {Intrusion} {Detection} in {Cloud} {Systems}},
	shorttitle = {{CIDS}},
	url = {https://ieeexplore.ieee.org/abstract/document/6209182},
	doi = {10.1109/ITNG.2012.94},
	urldate = {2025-06-17},
	booktitle = {{Ninth} {Int.} {Conf.} on {Information} {Technology} - {New} {Generations}},
	author = {Kholidy, Hisham A. and Baiardi, Fabrizio},
	month = apr,
	year = {2012},
	pages = {379--385},
}

@inproceedings{modi_novel_2013,
	title = {A novel hybrid-network intrusion detection system ({H}-{NIDS}) in cloud computing},
	url = {https://ieeexplore.ieee.org/abstract/document/6597201},
	doi = {10.1109/CICYBS.2013.6597201},
	urldate = {2025-06-17},
	booktitle = {{IEEE} {Symposium} on {Computational} {Intell.} in {Cyber} {Security} ({CICS})},
	author = {Modi, Chirag N. and Patel, Dhiren},
	month = apr,
	year = {2013},
	pages = {23--30},
}

@article{gul_distributed_2011,
	title = {Distributed {Cloud} {Intrusion} {Detection} {Model}},
	volume = {34},
	language = {en},
	journal = {Int. Journal of Advanced Science and Technology},
	author = {Gul, Irfan and Hussain, M},
	year = {2011},
}

@article{vieira_intrusion_2010,
	title = {Intrusion {Detection} for {Grid} and {Cloud} {Computing}},
	volume = {12},
	issn = {1941-045X},
	url = {https://ieeexplore.ieee.org/abstract/document/5232794},
	doi = {10.1109/MITP.2009.89},
	number = {4},
	urldate = {2025-06-17},
	journal = {IT Professional},
	author = {Vieira, Kleber and et al.},
	month = jul,
	year = {2010},
	pages = {38--43},
}

@incollection{xiang_taxonomy_2012,
	address = {Berlin, Heidelberg},
	title = {Taxonomy and {Proposed} {Architecture} of {Intrusion} {Detection} and {Prevention} {Systems} for {Cloud} {Computing}},
	volume = {7672},
	copyright = {http://www.springer.com/tdm},
	isbn = {978-3-642-35361-1 978-3-642-35362-8},
	url = {http://link.springer.com/10.1007/978-3-642-35362-8_33},
	language = {en},
	urldate = {2025-06-17},
	booktitle = {Cyberspace {Safety} and {Security}},
	publisher = {Springer Berlin Heidelberg},
	author = {Patel, Ahmed and et al.},
	editor = {Xiang, Yang and et al.},
	year = {2012},
	doi = {10.1007/978-3-642-35362-8_33},
	note = {Series Title: Lecture Notes in Comp. Science},
	pages = {441--458},
}

@article{javeed_federated_2024,
	title = {A federated learning-based zero trust intrusion detection system for {Internet} of {Things}},
	volume = {162},
	issn = {15708705},
	url = {https://linkinghub.elsevier.com/retrieve/pii/S1570870524001513},
	doi = {10.1016/j.adhoc.2024.103540},
	language = {en},
	urldate = {2025-06-13},
	journal = {Ad Hoc Networks},
	author = {Javeed, Danish and et al.},
	month = sep,
	year = {2024},
	pages = {103540},
}

@article{tlili_exhaustive_2024,
	title = {Exhaustive distributed intrusion detection system for {UAVs} attacks detection and security enforcement ({E}-{DIDS})},
	volume = {142},
	issn = {01674048},
	url = {https://linkinghub.elsevier.com/retrieve/pii/S0167404824001792},
	doi = {10.1016/j.cose.2024.103878},
	language = {en},
	urldate = {2025-06-13},
	journal = {Comp.s \& Security},
	author = {Tlili, Fadhila and et al.},
	month = jul,
	year = {2024},
	pages = {103878},
}

@article{sadia_intrusion_2024,
	title = {Intrusion {Detection} {System} for {Wireless} {Sensor} {Networks}: {A} {Machine} {Learning} {Based} {Approach}},
	volume = {12},
	copyright = {https://creativecommons.org/licenses/by-nc-nd/4.0/},
	issn = {2169-3536},
	shorttitle = {Intrusion {Detection} {System} for {Wireless} {Sensor} {Networks}},
	url = {https://ieeexplore.ieee.org/document/10477421/},
	doi = {10.1109/ACCESS.2024.3380014},
	urldate = {2025-06-13},
	journal = {IEEE Access},
	author = {Sadia, Halima and et al.},
	year = {2024},
	pages = {52565--52582},
}

@article{turukmane_m-multisvm_2024,
	title = {M-{MultiSVM}: {An} efficient feature selection assisted network intrusion detection system using machine learning},
	volume = {137},
	issn = {01674048},
	shorttitle = {M-{MultiSVM}},
	url = {https://linkinghub.elsevier.com/retrieve/pii/S0167404823004972},
	doi = {10.1016/j.cose.2023.103587},
	language = {en},
	urldate = {2025-06-13},
	journal = {Comp.s \& Security},
	author = {Turukmane, Anil V and Devendiran, Ramkumar},
	month = feb,
	year = {2024},
	pages = {103587},
}

@article{hnamte_dcnnbilstm_2023,
	title = {{DCNNBiLSTM}: {An} {Efficient} {Hybrid} {Deep} {Learning}-{Based} {Intrusion} {Detection} {System}},
	volume = {10},
	issn = {27725030},
	shorttitle = {{DCNNBiLSTM}},
	url = {https://linkinghub.elsevier.com/retrieve/pii/S2772503023000130},
	doi = {10.1016/j.teler.2023.100053},
	language = {en},
	urldate = {2025-06-13},
	journal = {Telematics and Informatics Reports},
	author = {Hnamte, Vanlalruata and Hussain, Jamal},
	month = jun,
	year = {2023},
	pages = {100053},
}

@article{lansky_deep_2021,
	title = {Deep {Learning}-{Based} {Intrusion} {Detection} {Systems}: {A} {Systematic} {Review}},
	volume = {9},
	copyright = {https://creativecommons.org/licenses/by/4.0/legalcode},
	issn = {2169-3536},
	shorttitle = {Deep {Learning}-{Based} {Intrusion} {Detection} {Systems}},
	url = {https://ieeexplore.ieee.org/document/9483916/},
	doi = {10.1109/ACCESS.2021.3097247},
	urldate = {2025-06-13},
	journal = {IEEE Access},
	author = {Lansky, Jan and et al.},
	year = {2021},
	pages = {101574--101599},
}

@article{abdulganiyu_systematic_2023,
	title = {A systematic literature review for network intrusion detection system ({IDS})},
	volume = {22},
	issn = {1615-5262, 1615-5270},
	url = {https://link.springer.com/10.1007/s10207-023-00682-2},
	doi = {10.1007/s10207-023-00682-2},
	language = {en},
	number = {5},
	urldate = {2025-06-13},
	journal = {Inter. Journal of Information Security},
	author = {Abdulganiyu, Oluwadamilare Harazeem and et al.},
	month = oct,
	year = {2023},
	pages = {1125--1162},
}

@article{jalil_hadi_real-time_2024,
	title = {Real-{Time} {Collaborative} {Intrusion} {Detection} {System} in {UAV} {Networks} {Using} {Deep} {Learning}},
	volume = {11},
	copyright = {https://ieeexplore.ieee.org/Xplorehelp/downloads/license-information/IEEE.html},
	issn = {2327-4662, 2372-2541},
	url = {https://ieeexplore.ieee.org/document/10594772/},
	doi = {10.1109/JIOT.2024.3426511},
	number = {20},
	urldate = {2025-06-13},
	journal = {IEEE Internet of Things Journal},
	author = {Jalil Hadi, Hassan and et al.},
	month = oct,
	year = {2024},
	pages = {33371--33391},
}

@article{alsaadi_adapting_2022,
	title = {An adapting soft computing model for intrusion detection system},
	volume = {38},
	issn = {0824-7935, 1467-8640},
	url = {https://onlinelibrary.wiley.com/doi/10.1111/coin.12433},
	doi = {10.1111/coin.12433},
	language = {en},
	number = {3},
	urldate = {2025-06-13},
	journal = {Computational Intell.},
	author = {Alsaadi, Husam Ibrahiem Husain and et al.},
	month = jun,
	year = {2022},
	pages = {855--875},
}

@inproceedings{fung_robust_2009,
	address = {New York, NY, USA},
	title = {Robust and scalable trust management for collaborative intrusion detection},
	isbn = {978-1-4244-3486-2},
	url = {http://ieeexplore.ieee.org/document/5188784/},
	doi = {10.1109/INM.2009.5188784},
	urldate = {2025-06-13},
	booktitle = {{IFIP}/{IEEE} {Inter.} {Symposium} on {Integrated} {Network} {Management}},
	publisher = {IEEE},
	author = {Fung, Carol J. and Zhang, Jie and Aib, Issam and Boutaba, Raouf},
	month = jun,
	year = {2009},
	pages = {33--40},
}

@article{min_cai_collaborative_2005,
	title = {Collaborative {Internet} {Worm} {Containment}},
	volume = {3},
	copyright = {https://ieeexplore.ieee.org/Xplorehelp/downloads/license-information/IEEE.html},
	issn = {1540-7993},
	url = {http://ieeexplore.ieee.org/document/1439499/},
	doi = {10.1109/MSP.2005.63},
	language = {en},
	number = {3},
	urldate = {2025-06-13},
	journal = {IEEE Security and Privacy Magazine},
	author = {{Min Cai} and et al.},
	month = may,
	year = {2005},
	pages = {25--33},
}

@inproceedings{janakiraman_indra_2003,
	address = {Linz, Austria},
	title = {Indra: a peer-to-peer approach to network intrusion detection and prevention},
	isbn = {978-0-7695-1963-0},
	shorttitle = {Indra},
	url = {http://ieeexplore.ieee.org/document/1231412/},
	doi = {10.1109/ENABL.2003.1231412},
	urldate = {2025-06-13},
	booktitle = {{WET} {ICE}. {Proce.}. {Twelfth} {IEEE} {Inter.} {Workshops} on {Enabling} {Technologies}: {Infrastructure} for {Collaborative} {Enterprises}, 2003.},
	publisher = {IEEE Comp. Soc},
	author = {Janakiraman, R. and et al.},
	year = {2003},
	pages = {226--231},
}

@inproceedings{zhou_evaluation_2007,
	address = {Munich, Germany},
	title = {Evaluation of a {Decentralized} {Architecture} for {Large} {Scale} {Collaborative} {Intrusion} {Detection}},
	isbn = {978-1-4244-0798-9 978-1-4244-0799-6},
	url = {http://ieeexplore.ieee.org/document/4258524/},
	doi = {10.1109/INM.2007.374772},
	urldate = {2025-06-13},
	booktitle = {10th {IFIP}/{IEEE} {Inter.} {Symposium} on {Integrated} {Network} {Management}},
	publisher = {IEEE},
	author = {Zhou, Chenfeng Vincent and et al.},
	month = may,
	year = {2007},
	pages = {80--89},
}

@article{martinez_beltran_decentralized_2023,
	title = {Decentralized {Federated} {Learning}: {Fundamentals}, {State} of the {Art}, {Frameworks}, {Trends}, and {Challenges}},
	volume = {25},
	issn = {1553-877X},
	shorttitle = {Decentralized {Federated} {Learning}},
	url = {https://ieeexplore.ieee.org/abstract/document/10251949},
	doi = {10.1109/COMST.2023.3315746},
	number = {4},
	urldate = {2025-06-12},
	journal = {IEEE Communications Surveys \& Tutorials},
	author = {Martínez Beltrán, Enrique Tomás and et al.},
	year = {2023},
	pages = {2983--3013},
}

@inproceedings{wink_approach_2021,
	title = {An {Approach} for {Peer}-to-{Peer} {Federated} {Learning}},
	url = {https://ieeexplore.ieee.org/abstract/document/9502443},
	doi = {10.1109/DSN-W52860.2021.00034},
	urldate = {2025-06-12},
	booktitle = {2021 51st {Annual} {IEEE}/{IFIP} {Inter.} {Conf.} on {Dependable} {Systems} and {Networks} {Workshops} ({DSN}-{W})},
	author = {Wink, Tobias and Nochta, Zoltan},
	month = jun,
	year = {2021},
	note = {ISSN: 2325-6664},
	pages = {150--157},
}

@article{vasilomanolakis_taxonomy_2015,
	title = {Taxonomy and {Survey} of {Collaborative} {Intrusion} {Detection}},
	volume = {47},
	issn = {0360-0300},
	url = {https://dl.acm.org/doi/10.1145/2716260},
	doi = {10.1145/2716260},
	number = {4},
	urldate = {2025-06-12},
	journal = {ACM Comp. Surv.},
	author = {Vasilomanolakis, Emmanouil and et al.},
	year = {2015},
	pages = {55:1--55:33},
}

@article{salehie_self-adaptive_2009,
	title = {Self-adaptive software: {Landscape} and research challenges},
	volume = {4},
	issn = {1556-4665},
	shorttitle = {Self-adaptive software},
	url = {https://dl.acm.org/doi/10.1145/1516533.1516538},
	doi = {10.1145/1516533.1516538},
	number = {2},
	urldate = {2025-06-12},
	journal = {ACM Trans. Auton. Adapt. Syst.},
	author = {Salehie, Mazeiar and Tahvildari, Ladan},
	month = may,
	year = {2009},
	pages = {14:1--14:42},
}

@article{zhou_survey_2010,
	title = {A survey of coordinated attacks and collaborative intrusion detection},
	volume = {29},
	issn = {0167-4048},
	url = {https://www.sciencedirect.com/science/article/pii/S016740480900073X},
	doi = {10.1016/j.cose.2009.06.008},
	number = {1},
	urldate = {2025-06-12},
	journal = {Comp.s \& Security},
	author = {Zhou, Chenfeng Vincent and et al.},
	month = feb,
	year = {2010},
	pages = {124--140},
}

@inproceedings{hardegen_hierarchical_2021,
	title = {A {Hierarchical} {Architecture} and {Probabilistic} {Strategy} for {Collaborative} {Intrusion} {Detection}},
	url = {https://ieeexplore.ieee.org/abstract/document/9705027},
	doi = {10.1109/CNS53000.2021.9705027},
	urldate = {2025-05-16},
	booktitle = {{IEEE} {Conf.} on {Communications} and {Network} {Security} ({CNS})},
	author = {Hardegen, Christoph and et al.},
	month = oct,
	year = {2021},
	pages = {128--136},
}

@inproceedings{cuppens_alert_2002,
	title = {Alert correlation in a cooperative intrusion detection framework},
	url = {https://ieeexplore.ieee.org/document/1004372},
	doi = {10.1109/SECPRI.2002.1004372},
	urldate = {2025-05-16},
	booktitle = {Proce. {IEEE} {Symposium} on {Security} and {Privacy}},
	author = {Cuppens, F. and Miege, A.},
	month = may,
	year = {2002},
	pages = {202--215},
}

@inproceedings{duma_trust-aware_2006,
	title = {A {Trust}-{Aware}, {P2P}-{Based} {Overlay} for {Intrusion} {Detection}},
	url = {https://ieeexplore.ieee.org/abstract/document/1698432},
	doi = {10.1109/DEXA.2006.21},
	urldate = {2025-05-16},
	booktitle = {17th {Inter.} {Workshop} on {Database} and {Expert} {Systems} {Applications} ({DEXA}'06)},
	author = {Duma, C. and et al.},
	month = sep,
	year = {2006},
	pages = {692--697},
}

@book{debar_rfc_2007,
	address = {USA},
	title = {{RFC} 4765: {The} {Intrusion} {Detection} {Message} {Exchange} {Format} ({IDMEF})},
	shorttitle = {{RFC} 4765},
	publisher = {RFC Editor},
	author = {Debar, H. and et al.},
	month = feb,
	year = {2007},
}

@article{morais_distributed_2014,
	title = {A {Distributed} and {Collaborative} {Intrusion} {Detection} {Architecture} for {Wireless} {Mesh} {Networks}},
	volume = {19},
	issn = {1572-8153},
	url = {https://doi.org/10.1007/s11036-013-0457-8},
	doi = {10.1007/s11036-013-0457-8},
	language = {en},
	number = {1},
	urldate = {2025-05-16},
	journal = {Mobile Networks and Applications},
	author = {Morais, Anderson and Cavalli, Ana},
	month = feb,
	year = {2014},
	pages = {101--120},
}

@article{dasgupta_cids_2005,
	title = {{CIDS}: {An} agent-based intrusion detection system},
	volume = {24},
	issn = {0167-4048},
	shorttitle = {{CIDS}},
	url = {https://www.sciencedirect.com/science/article/pii/S0167404805000179},
	doi = {10.1016/j.cose.2005.01.004},
	number = {5},
	urldate = {2025-05-16},
	journal = {Comp.s \& Security},
	author = {Dasgupta, D. and et al.},
	month = aug,
	year = {2005},
	pages = {387--398},
}

@inproceedings{ezelu_membership_2023,
	title = {Membership {Management} in {Collaborative} {Intrusion} {Detection} {Systems}},
	url = {https://ieeexplore.ieee.org/document/10073989/},
	doi = {10.1109/ICNC57223.2023.10073989},
	urldate = {2025-05-16},
	booktitle = {{Int.} {Conf.} on {Computing}, {Networking} and {Communications} ({ICNC})},
	author = {Ezelu, Chukwuebuka and et al.},
	month = feb,
	year = {2023},
	pages = {47--53},
}

@article{landauer_aminer_2023,
	title = {{AMiner}: {A} {Modular} {Log} {Data} {Analysis} {Pipeline} for {Anomaly}-based {Intrusion} {Detection}},
	volume = {4},
	shorttitle = {{AMiner}},
	url = {https://dl.acm.org/doi/10.1145/3567675},
	doi = {10.1145/3567675},
	number = {1},
	urldate = {2025-02-10},
	journal = {Digital Threats},
	author = {Landauer, Max and et al.},
	year = {2023},
	pages = {12:1--12:16},
}

@inproceedings{gurung_cids_2022,
	title = {{CIDS}: {Collaborative} {Intrusion} {Detection} {System} using {Blockchain} {Technology}},
	shorttitle = {{CIDS}},
	url = {https://ieeexplore.ieee.org/abstract/document/9850331},
	doi = {10.1109/CSR54599.2022.9850331},
	urldate = {2025-05-16},
	booktitle = {{IEEE} {Int.} {Conf.} on {Cyber} {Security} and {Resilience} ({CSR})},
	author = {Gurung, Gopal and et al.},
	month = jul,
	year = {2022},
	pages = {125--130},
}

@article{javed_prism_2023,
	title = {{PRISM}: {A} {Hierarchical} {Intrusion} {Detection} {Architecture} for {Large}-{Scale} {Cyber} {Networks}},
	volume = {20},
	issn = {1941-0018},
	shorttitle = {{PRISM}},
	url = {https://ieeexplore.ieee.org/document/10027167/},
	doi = {10.1109/TDSC.2023.3240315},
	number = {6},
	urldate = {2025-05-16},
	journal = {IEEE Transactions on Dependable and Secure Computing},
	author = {Javed, Yahya and et al.},
	month = nov,
	year = {2023},
	pages = {5070--5086},
}

@inproceedings{vasilomanolakis_skipmon_2015,
	title = {{SkipMon}: {A} locality-aware {Collaborative} {Intrusion} {Detection} {System}},
	shorttitle = {{SkipMon}},
	url = {https://ieeexplore.ieee.org/abstract/document/7410282},
	doi = {10.1109/PCCC.2015.7410282},
	urldate = {2025-05-16},
	booktitle = {{IEEE} 34th {Int.} {Performance} {Computing} and {Communications} {Conf.} ({IPCCC})},
	author = {Vasilomanolakis, Emmanouil and et al.},
	month = dec,
	year = {2015},
	pages = {1--8},
}

@inproceedings{locasto_towards_2005,
	title = {Towards collaborative security and {P2P} intrusion detection},
	url = {https://ieeexplore.ieee.org/document/1495971},
	doi = {10.1109/IAW.2005.1495971},
	urldate = {2025-05-16},
	booktitle = {Proce. from the {Sixth} {Annual} {IEEE} {SMC} {Information} {Assurance} {Workshop}},
	author = {Locasto, M.E. and et al.},
	month = jun,
	year = {2005},
	pages = {333--339},
}

@inproceedings{wu_collaborative_2003,
	title = {Collaborative intrusion detection system ({CIDS}): a framework for accurate and efficient {IDS}},
	shorttitle = {Collaborative intrusion detection system ({CIDS})},
	url = {https://ieeexplore.ieee.org/abstract/document/1254328},
	doi = {10.1109/CSAC.2003.1254328},
	urldate = {2025-05-19},
	booktitle = {19th {Annual} {Comp.} {Security} {Applications} {Conf.}, 2003. {Proce.}.},
	author = {Wu, Yu-Sung and et al.},
	month = dec,
	year = {2003},
	pages = {234--244},
}

@article{arshad_colide_2019,
	title = {{COLIDE}: a collaborative intrusion detection framework for {Internet} of {Things}},
	volume = {8},
	issn = {2047-4954, 2047-4962},
	shorttitle = {{COLIDE}},
	url = {https://ietresearch.onlinelibrary.wiley.com/doi/10.1049/iet-net.2018.5036},
	doi = {10.1049/iet-net.2018.5036},
	language = {en},
	number = {1},
	urldate = {2025-05-19},
	journal = {IET Net.},
	author = {Arshad, Junaid and et al.},
	month = jan,
	year = {2019},
	pages = {3--14},
}

@inproceedings{dautov_latency-aware_2024,
	title = {Latency-{Aware} {Node} {Selection} in {Federated} {Learning}},
	url = {https://ieeexplore.ieee.org/document/10619860},
	doi = {10.23919/IFIPNetworking62109.2024.10619860},
	urldate = {2025-06-05},
	booktitle = {{IFIP} {Networking} {Conf.} ({IFIP} {Networking})},
	author = {Dautov, Rustem and et al.},
	month = jun,
	year = {2024},
	note = {ISSN: 1861-2288},
	pages = {598--600},
}

@article{landauer2026attackmate,
  title={AttackMate: Realistic Emulation and Automation of Cyber Attack Scenarios Across the Kill Chain},
  author={Landauer, Max and et al.},
  journal={arXiv preprint arXiv:2601.14108},
  year={2026}
}

@dataset{hotwagner_2026_18482763,
  author       = {Hotwagner, Wolfgang and et al.},
  title        = {AttackMate Robotdog Log Dataset},
  month        = feb,
  year         = 2026,
  publisher    = {Zenodo},
  doi          = {10.5281/zenodo.18482763},
  url          = {https://doi.org/10.5281/zenodo.18482763},
}

\end{document}